\documentclass[twocolumn,10pt]{revtex4-1}
\usepackage{amsmath,amssymb}
\usepackage{graphicx}
\usepackage{caption}
\usepackage{subcaption}
\begin{document}

\title{Gravitational Tensor-Monopole Moment of Hydrogen Atom To Order ${\cal O}(\alpha)$}

\author{Xiangdong Ji}
\email{xji@umd.edu}

\affiliation{Maryland Center for Fundamental Physics,
Department of Physics, University of Maryland,
College Park, 20742, USA}

\author{Yizhuang Liu}
\email{yizhuang.liu@uj.edu.pl}
\affiliation{Institute of Theoretical Physics, Jagiellonian University, 30-348 Kraków, Poland}

\date{\today}
\begin{abstract}
We calculate the gravitational tensor-monopole moment of the momentum-current density $T^{ij}$ in the ground state of the hydrogen atom to
order ${\cal O}(\alpha)$ in quantum electrodynamics (QED). The result is
\begin{align}
\tau_H/\tau_0 - 1 = \frac{4\alpha}{3\pi}\left(\ln\alpha^2-0.028\right)
\nonumber
\end{align}
where $\tau_0= \hbar^2/4m_e$ is the leading-order moment.
The physics of the next-to-leading-order correction is similar
to that of the famous Lamb shift for energy levels.
\end{abstract}

\maketitle

\section{introduction}
The energy-momentum-tensor (EMT) form factors for hadrons in quantum chromodynamics (QCD) are physical quantities that can be measured through deeply-virtual Compton scattering~\cite{Ji:1996ek,Ji:1996nm,HERMES:2001bob,CLAS:2001wjj,ZEUS:2003pwh,H1:2005gdw} and similar processes discussed in Refs. \cite{Radyushkin:1996ru,Kharzeev:1998bz,Berger:2001xd,Frankfurt:2002ka,Guidal:2002kt,Hatta:2018ina,Guo:2021ibg,Qiu:2022bpq}. On the one hand, they are related to the perturbation of space-time induced by the hadrons; on the other hand, these form factors can also be used to characterize internal structure, such as mass, spin and momentum-current
distributions. The physical interpretation of the momentum-current (MC) ($T^{ij}(\vec{q})$) form factor $C(q)$ (or $D(q)$) in the static limit sometimes differs across the literature. They are often used in a way that assumes similarity to macroscopic fluid. For example, various components of the momentum current has been assigned the meaning of ``pressure'' and ``shear pressure" in Refs.~\cite{Polyakov:2002yz,Polyakov:2018zvc,Burkert:2018bqq,Shanahan:2018nnv}, and the ``mechanical stability" condition~\cite{Polyakov:2018zvc,Lorce:2018egm} further implies that the $D$-term form factor~\cite{Polyakov:1999gs} is negative at $q=0$.  In our previous paper~\cite{Ji:2021mfb}, we suggested interpreting the form factors in terms of gravitational multipoles according to their
role in generating static gravity nearby. In particular, the gravitational tensor-monopole moment $T0$ of the momentum current is related to the $C(q^2=0)$ form factor~\cite{Ji:1996ek}.  As a concrete example, we have calculated the tensor-monopole moment $\tau_H$ of hydrogen-like atom in quantum electrodynamics (QED) to leading order in the fine structure constant
$\alpha=e^2/4\pi$, and the sign is opposite of the ``mechanical stability" condition, showing the concept has little relevance in quantum mechanical systems.

As a reminder, we recall the definition of the gravitational tensor-monopole moment for static momentum-current distribution, $T^{ij}(\vec{x})$. According to our previous paper~\cite{Ji:2021mfb}, the {\it tensor monopole} $T0$ is defined as:
 \begin{align}
     T^{(0)} =\frac{1}{5}\int d^3\vec{x}T_{ij}(\vec{x})\left(x_ix_j-\frac{\delta_{ij}}{3}x^2\right)  \ .
 \end{align}
Using the conservation law or transverse condition $\partial_i T^{ij}=0$, one can show that it is related to {\it scalar momentum-current radius} ,
\begin{align}
    T^{(0)} =-\frac{1}{6}\int d^3\vec{x} x^2 T_{ii}(\vec{x}) \ ,
\end{align}
where $T^{ii}(\vec{x})$ is proportional to the so-called pressure $p(r)$ in the other literature following a continuous medium~\cite{Polyakov:2002yz,Polyakov:2018zvc,Burkert:2018bqq,Shanahan:2018nnv}, although
the negative $p(r)$ demands an interpretation that deviates from the standard thermodynamics.
We choose to normalize the
tensor-monopole moment of a system as
\begin{equation}
    \tau =-T^{(0)}/2 \ ,
\end{equation}
which relates to the ``$D$-term'' $D(0)$~\cite{Polyakov:1999gs} as $\tau=\frac{D(0)}{4M}$ where $M$ is the total mass of the system.

In this paper we will consider the ground state of hydrogen atom in the infinitely-heavy proton limit, where the position of the proton is fixed at origin, while the electron and photon part is treated in the background field formalism as in Ref.~\cite{Weinberg:1995mt}. In particular, the ground state $|0\rangle \equiv |0\rangle_H$ of the hydrogen atom will not be translation  invariant, but describing a static spherical symmetric energy-momentum distribution around the origin. Therefore, the expectation value $\langle T^{ij}\rangle(\vec{x}) \equiv \langle 0|T^{ij}(\vec{x})|0\rangle$ of the MC density operator $T^{ij}(\vec{x})$ in the ground state $|0\rangle$ can be viewed as the MC density distribution of a classical system, from which the tensor-monopole moment can be defined as above.  Due to the spherical symmetry of the ground state, in momentum space one has
\begin{align}
    \langle T^{ij}\rangle (\vec{q})=(q^iq^j-\delta^{ij}q^2)\frac{C_H(q)}{m_e} \ ,
\end{align}
where $\vec{q}$ is the 3-momentum transfer taken place at the insertion of $T^{ij}$.
After a simple calculation, one can show
\begin{equation} \label{eq:deftau}
         \tau = \frac{C_H(q=0)}{m_e} \ ,
\end{equation}
where $m_e$ is the electron mass.

In this paper, following up Ref.~\cite{Ji:2021mfb}, we calculate the $\tau_H$ for the hydrogen atom
to electromagnetic order ${\cal O}(\alpha)$ by including leading-order radiative corrections. The calculation is not entirely trivial since the $\tau_H$ is infrared (IR) sensitive and exhibit logarithmic IR divergences in case of a single electron~\cite{Berends:1975ah,Milton:1977je,Donoghue:2001qc}. For hydrogen-like atom, the IR divergence is regulated by the binding energy differences of order $\alpha^2 m_e$ between the ground state and excited states, but the ultraviolet contributions remain the same as the free QED. The natural scale separation $m_e\gg \alpha m_e \gg \alpha^2m_e$ and the fact that the physics of the different energy scales decouple through logarithms allows the simplification of the calculation by first working in the non-relativistic versions of QED (NRQED)~\cite{Caswell:1985ui,Labelle:1996en,Pineda:1997bj}, and then match to full-QED,  in a way similar to the simplified calculation of the famous Lamb shift~\cite{Labelle:1996uc,Pineda:1997ie}.

The organization of the paper is as follows. In Section II, We first review the NRQED Lagrangian and introduce the effective EMT operator. To match to the EMT in QED, local counter-terms are required, which will be calculated in the following section.  We provide a short review the calculation at leading order, emphasizing the Coulomb contributions. We then present all the relevant diagrams at one-loop level, and show by explicit power-counting that only the photonic diagram will contribute to $\tau_H$ at order $\frac{\alpha}{m_e}$. In section III, we perform the matching of EMT in NRQED to order $\frac{\alpha}{m_e}$.  In section IV, we perform the calculation for the bound state. Combining the results of the two sections, we obtains the final result in Eq.~(\ref{eq:finalre}) and evaluate the sum numerically. Section V concludes the paper.

\section{Review of NRQED and overall strategy of the calculation}

In this section we introduce the EMT in NRQED and set up the overall strategy of calculating
the scalar form factor $C$ in hydrogen atom.  We review the leading-order contribution to the tensor-monopole moment $\tau_H$ of the momentum current. We show that by combining the fermionic contribution, the Coulomb self-interactions and the Coulomb interference between the electron and the proton, all the Coulomb tails get removed and the resulting monopole moment $\tau$ is equal to the basic unit $\tau_0=\hbar^2/4m_e$ and {\it positive}. This example shows that the sign of the $D$-term has little to do with the ``mechanical stability''.

We consider the bound state in quantum electrodynamics (QED) between two types of fermions, the standard negative charged electron with mass $m_e$ and positive charged ``proton'' with mass $M$. At energy scale much smaller than the proton mass $M$, the proton can be approximated by an infinitely-heavy static source fixed at $\vec{x}_p=0$ ,which sources the background electric potential
\begin{align}
    \nabla^2 V_p(r)=-e\delta^3(\vec{x}) \ ,
\end{align}
or $V_p=e/(4\pi r)$ ($r=|\vec{x}|$ and $e$ is the proton charge and positive). The system therefore reduces to a single electron moving in the presence of the background field $V_p$. In principle, it can be described by the dressed-Dirac theory discussed in Ref.~\cite{Weinberg:1995mt} where a complete set of solutions to the Dirac equation are being used to define the free theory, upon which radiative corrections can be added consistently. This dressed-Dirac theory has the same short distance behavior as the free QED and can be renormalized using the same renormalization constants $Z_1=Z_2$ and $Z_3$ as the free QED. This will be the underline ``first principle'' theory of this paper.

\begin{figure}[t]
{%
  \includegraphics[height=2cm]{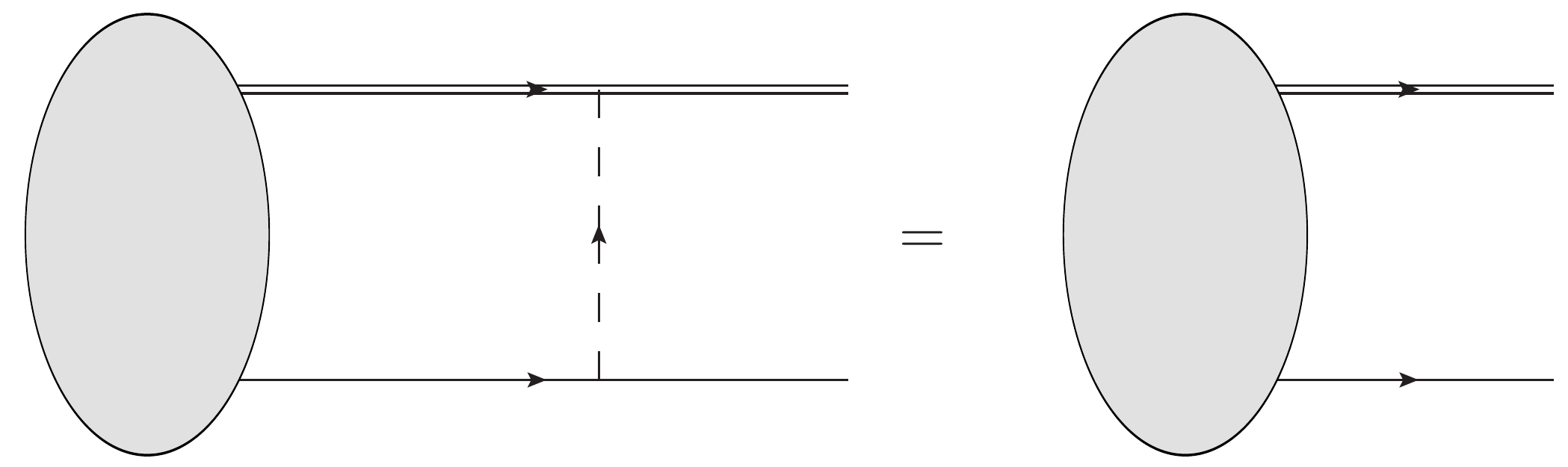}
}
\caption{The Bethe-Salpeter equation for the wave function $\Phi$ denoted by the oval blob. Double line represents propagator of proton field and single line represents the electron propagator. The dashed line represents the exchange of a Coulomb photon. }
\label{fig:Bethe}
\end{figure}

Equivalently, the first principle theory can also be treated using the Bethe-Salpeter equation approach, where the quantum nature of the proton can be preserved. See Fig.~\ref{fig:Bethe} for a depiction of the Bethe-Salpeter equation approach. Although equivalent in the infinite heavy proton limit, in this paper we will use the background field approach without mentioning otherwise.

For small fine structure constant $\alpha$, bound-states and low-energy excited states of the Dirac equation are essentially non-relativistic, and to leading order in $\alpha$ reduces to the
standard Schrodinger equation
\begin{align}\label{eq:schor}
    \left(-\frac{1}{2m_e}\nabla^2-eV_p(r)\right)\psi(\vec{x})=E\psi(\vec{x}) \ .
\end{align}
The bound-state is characterized by two scales, the binding energy $\alpha^2m_e$ and the inverse Bohr radius $\alpha m_e$.  One also needs the complete set of energy eigenfunctions of the above Schrodinger equation $\psi_{M}(\vec{x})$ with energy $E_M$ and the normalization condition
\begin{align}\label{eq:complet}
\sum_{M}\psi_{M}(\vec{x})\psi^{\dagger}_M(\vec{y})=\delta^{3}(\vec{x}-\vec{y}) \ .
\end{align}
It will always be understood that the set of wave functions $\psi_M$ contains both discrete and continuum spectrum. As we will show later, the contribution of the continuum spectrum is not negligible.

\subsection{Basics of NRQED}
The approach which starts from the Bethe-Salpeter equation or the fully-dressed Dirac theory in the background field as explained in \cite{Weinberg:1995mt},  is simple to understand but hard to use.  This is mainly due to the complicate form of the Dirac-Coulomb propagator. For bound states with large atomic number $Z$ where relativistic effect is large, one must calculate radiative corrections numerically with Dirac-Coulomb propagator. For $Z=1$, however, the non-relativistic nature of the bound state and the emergence of the scale separation $\alpha^2m_e \ll \alpha m_e \ll m_e$ allows dramatic simplification after performing ``twist expansion'' or non-relativistic expansion in the soft scales.

The modern way to organize this expansion is through the effective field theory, more precisely, the non-relativistic reduction of QED (NRQED)~\cite{Caswell:1985ui,Labelle:1996en,Pineda:1997bj}. In NRQED, all the effective fields $\Psi$ and $A^i$ contains momentum scale comparable of smaller than $\alpha m_e$, while ultra-violet contributions are integrated out into local operators at the Lagrangian level order by order in $\frac{1}{m_e}$ , with $\alpha$-dependent {\it matching coefficients } in order to match the radiative corrections to the full-theory when expanded to the same order in $\frac{1}{m_e}$. Although not proven, it is widely believed such matching can be performed consistently to all orders in $\frac{1}{m_e}$ and $\alpha$.

For our purpose, namely, calculating the tensor-monopole moment to order $\alpha$, one only needs the Lagrangian of NRQED to order $\frac{1}{m_e}$:
\begin{align}\label{eq:LNRQED}
    {\cal L}_{\rm NRQED}=&\Psi^{\dagger}(iD_0+\frac{D_iD_i}{2m_e})\Psi-c_F\frac{e}{2m_e}\Psi^{\dagger}\vec{\sigma}\cdot \vec{B}\Psi \nonumber \\
   &-\frac{1}{4}F_{\mu\nu}F^{\mu\nu}+{\cal O}(\frac{1}{m_e^2}) \ ,
\end{align}
where $D^0=\partial^0-ieA^0-ieV_p$ contains the background field $V_p$ and $D^i=\partial^i-ieA^i$. The $F^{\mu\nu}=\partial^{\mu}A^{\nu}-\partial^{\nu}A^{\mu}$ is the standard field strength,  $B^{i}=\frac{1}{2}\epsilon^{ijk}F^{jk}$ is the magnetic field and $c_F=1+\frac{\alpha}{2\pi}$ is the matching constants that is required to match the spin part of the vertex function in the effective theory to the full QED (to this order it is just the famous anomalous magnetic moment). Furthermore, since we are only interested in the spin-independent part of the EMT form factor to order $\frac{\alpha}{m_e}$, the spin contribution can be neglected and one needs only
\begin{align}\label{eq:NRLag}
 {\cal L}_{\rm NRQED}=\Psi^{\dagger}(iD_0+\frac{D_iD_i}{2m_e})\Psi-\frac{1}{4}F_{\mu\nu}F^{\mu\nu}  \ ,
\end{align}
which will be used throughout the paper.

With the complete set of energy wave functions given in Eq.~(\ref{eq:complet}) serving as the fundamental basis for the free-electron field,  the above Lagrangian can be quantized in the Coulomb gauge as usual, where $A^0$
\begin{align}
    A^0\equiv V_e=\frac{1}{\nabla^2}e\Psi^{\dagger}\Psi \ ,
\end{align}
are being solved in terms of the electron field explicitly, and $A^i$ contains only the transverse part, $\nabla \cdot \vec{A}=0$. In this gauge, the electric field can be separated into longitudinal and transverse (radiative) parts as
\begin{align}
    \vec{E}=\vec{E}_{\parallel}+\vec{E}_{\perp} \ , \\
    \vec{E}_{\parallel}=-\nabla V_e \ , \\
    \vec{E}_{\perp}=-\partial^0 \vec{A} \ ,
\end{align}
and for the magnetic field there is only $\vec{B}_{\perp}$.  Although not appearing in the lagrangian, the electric field of the proton reads
\begin{align}
    \vec{E}_p=-\nabla V_p=\frac{e\vec{x}}{4\pi |\vec{x}|^3} \ ,
\end{align}
which will enter in the total EMT.

For the above fields in NRQED, their $\alpha$-counting rules are as follows.
For the electron field, the momentum scale is always $\alpha m$, which implies
\begin{align}\label{eq:powerphi}
\Psi^{\dagger}\Psi \sim \alpha^{3} \ .
\end{align}
For the bound states, the above Lagrangian contains contributions from both the soft photon with $|\vec{k}|={\cal O}(\alpha m_e)$ and the ultra-soft photon with $|\vec{k}|={\cal O}(\alpha^2 m_e)$. For soft radiative photon with $|\vec{k}|={\cal O}(\alpha m_e)$, one has
\begin{align}\label{eq:powerrs}
   |e\vec{E}_{\perp}|_{s},  |e\vec{B}_{\perp}|_{s} \sim \alpha^4 \ ,
\end{align}
while for ultra-soft radiative photon one has
\begin{align}\label{eq:powerrus}
    |e\vec{E}_{\perp}|_{us},  |e\vec{B}_{\perp}|_{us}  \sim \alpha^{6} \ .
\end{align}
For Coulomb photon which is soft, one has
\begin{align}\label{eq:powercs}
     |e\vec{E}_{\parallel}|_s \sim \alpha^3 \ .
\end{align}
When coupled to ultra-soft radiative photon, the Coulomb photon can also become ultra-soft. The power-counting in this case is tricky. Indeed, in coordinate state one has
\begin{align}
\langle N|\vec{E}_{\parallel}(\vec{x})|M\rangle=\int \frac{d^3\vec{k}}{(2\pi)^3 }\frac{ie\vec{k}}{|\vec{k}|^2}e^{i\vec{k}\cdot \vec{x}}(\psi^{\dagger}_M \psi_N)(\vec{k}) \ ,
\end{align}
therefore, for $|\vec{k}|={\cal O}(\alpha^2 m_e)$ the naive power-counting reads $|e\vec{E}_{\parallel}| \sim \alpha^{5}$. However, when $M \ne N$, in the matrix element one must Taylor expanding to next-leading order,
\begin{align}\label{eq:Ecspecial}
(\psi^{\dagger}_M \psi_N)(\vec{k})=-i\vec{k}\cdot \langle M|\vec{x} |N\rangle \sim \alpha \ ,
\end{align}
leading to one more power of $\alpha$, therefore in this situation the power-counting rule for ultra-soft Coulomb photon will be
\begin{align}\label{eq:powercus}
|e\vec{E}_{\parallel}|_{us} \sim \alpha^{6} \ ,
\end{align}
the same as ultra-soft radiative photon.

To regulate the UV divergences, one must specify the UV regulator, which we chose to be the standard dimensional regulator with $D=3-2\epsilon$. It has been applied in NRQED to calculate the famous Lamb~\cite{Pineda:1997ie},  ${\cal O}(\alpha^5m_e)$~\cite{Pineda:1998kn} and ${\cal O}(\alpha^6m_e)$~\cite{Czarnecki:1999mw} corrections to positronium spectrum. This means, in the intermediate steps of calculation, one must use the wave functions and matrix elements in $D=3-2\epsilon$ dimensions. For example, the Schrodinger equation becomes
\begin{align}\label{eq:schorD}
    \left(-\frac{1}{2m_e}\nabla^2-eV_{p,D}(r)\right)\psi(\vec{x})=E\psi(\vec{x}) \ ,
\end{align}
with $D$-dependent potential
\begin{align}
    V_{p,D}(r)=\mu^{2\epsilon}\int \frac{d^D{\vec{p}}}{(2\pi)^D|\vec{p}|^2}e^{i\vec{p}\cdot\vec{x}} \ ,
\end{align}
and the normalization condition
\begin{align}
\int d^D\vec{x} \psi^{\dagger}_M(\vec{x})\psi_N(\vec{x})=\delta_{MN} \ .
\end{align}
In terms of them, one can form the matrix elements
\begin{align}\label{eq:matrix}
\rho_{NM}(\vec{k})=\int d^D\vec{x} \psi^{\dagger}_N(\vec{x})\psi_{M}(\vec{x})e^{-i\vec{k}\cdot\vec{x}} \ , \\
\vec{v}_{NM}(\vec{k})=\int d^D\vec{x} \psi^{\dagger}_N(\vec{x})\frac{-i\vec{\nabla}}{m_e}\psi_{M}(\vec{x})e^{-i\vec{k}\cdot\vec{x}} \ ,
\end{align}
which are $D$-dependent generically, and will appear in the bound-state calculation. However, we will show that at the final stage of the calculation, by using the sum rule
\begin{align}\label{eq:sumrule}
    \sum_{M}\frac{2\vec{v}_{MN}\cdot \vec{v}_{NM}}{D(E_M-E_N)} =\frac{1}{m_e} \ ,
\end{align}
coefficients of the $\frac{1}{\epsilon}$ poles are {\it $D$-independent} universal constants, while all other finite terms can be safely set to the value in $D=3$ without causing trouble.

\begin{figure}[htb]	
\begin{subfigure}{.2\textwidth}
		\centering
		\includegraphics[width=\linewidth]{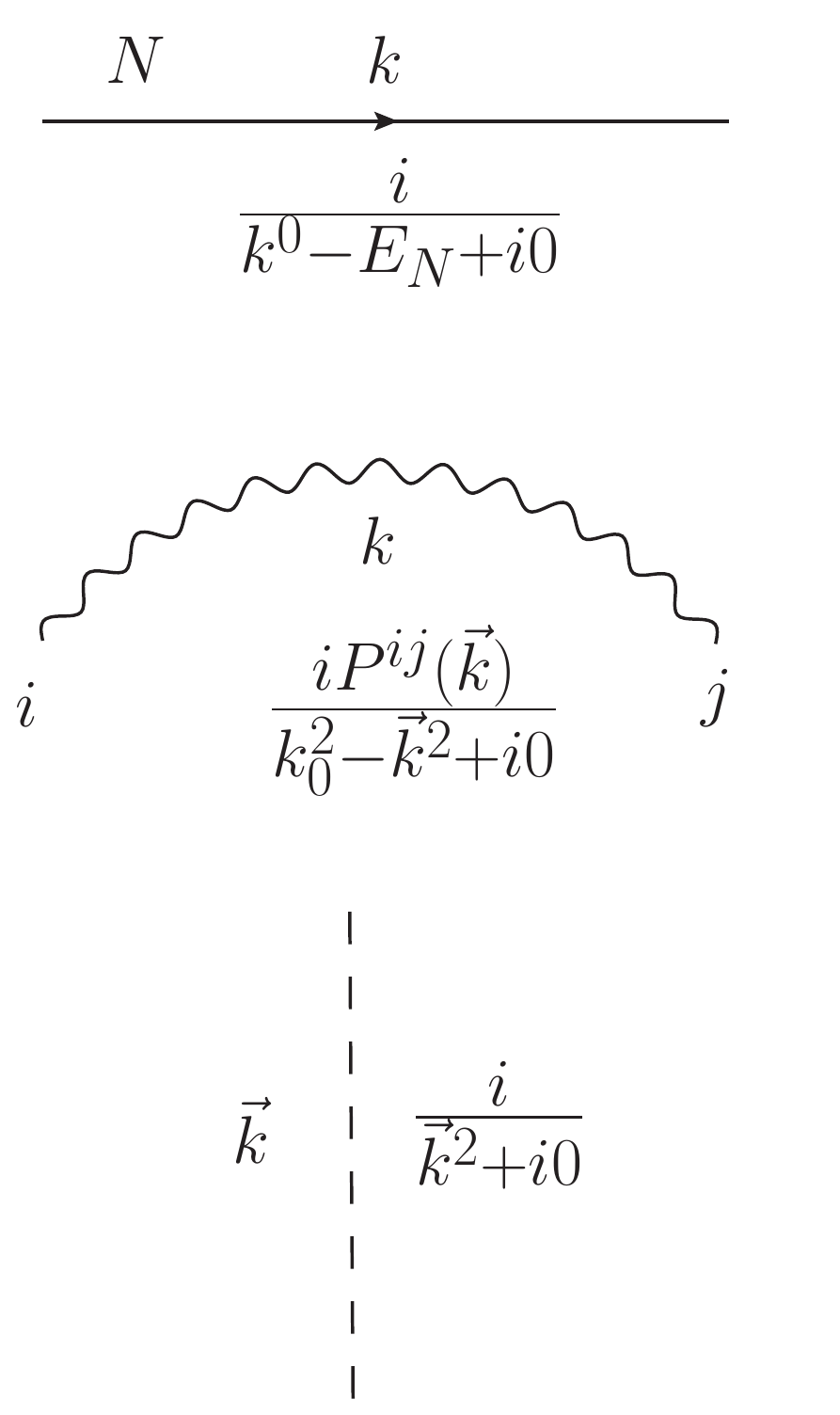}
		\caption{}
		\label{fig:rulecovpro}
	\end{subfigure}
  \begin{subfigure}{.13\textwidth}
		\centering
		\includegraphics[width=\linewidth]{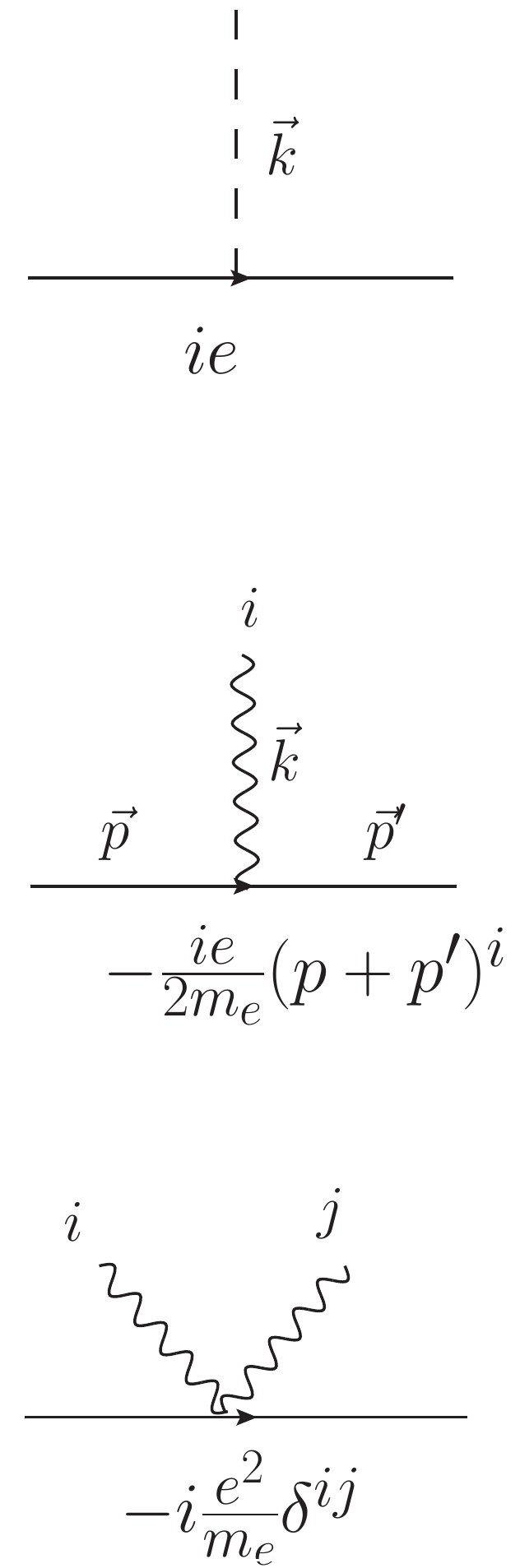}
		\caption{}
		\label{fig:rulecpvertex}
	\end{subfigure}
\caption{The Feynman rules of the NRQED Lagrangian Eq.~(\ref{eq:NRLag}). In Fig.~\ref{fig:rulecovpro}, we show the propagator for electron, radiative photon and Coulomb photon in covariant perturbation theory. In  Fig.~\ref{fig:rulecpvertex}, we show all the interaction vertices. To obtain the rule in old-fashioned perturbation theory, simply remove an $-i$ from the non-instantaneous triple electron-photon vertex and the seagull vertex. }
		\label{fig:rules}
\end{figure}

After introducing the Lagrangian and the UV regulator, we now list the Feynman-rules of the covariant perturbation theory in Fig.~\ref{fig:rules}. The radiative photon polarization
sum is
\begin{equation}
    P^{ij}(\vec{k}) = \delta^{ij} - \frac{k^ik^j}{|\vec{k}|^2} \ .
\end{equation}
All the interaction vertices are represented in plan-wave basis, in the energy eigen-basis they are replaced by the matrix-elements as in Eq.~(\ref{eq:matrix}). Fig.~\ref{fig:rulecpvertex} contains the standard instantaneous Coulomb vertices, the triple electron-photon vertices and the seagull vertex due to the $-\frac{e^2}{2m_e^2}A^iA^i\Psi^{\dagger}\Psi$ term in the Lagrangian. For non-relativistic system, sometimes it is convenient to use the old-fashioned perturbation theory as well, which can be obtained by integrating out the $k^0$ first and consists of matrix elements between free states followed by energy denominators. The matrix-elements can still be obtained from Fig.\ref{fig:rulecpvertex} by removing factor of $-i$'s for the non-instantaneous triple photon-electron vertex and the seagull vertex.

It is possible to separate the ultra-soft contribution to define a new effective theory, the potential-NRQED(pNRQED)~\cite{Pineda:1997bj,Pineda:1997ie}, where further simplifications for ultra-soft photons are performed, namely, expanding all the $e^{-i\vec{k}\cdot \vec{x}}$ in the form-factors in Eq.~(\ref{eq:matrix}). The contribution from the soft photons, however, can be calculated largely in the free-NRQED by expanding the electron propagators in the background field. Provided one uses the correct rules in the corresponding region, as we do here,
it is not necessary to introduce the pNRQED Lagrangian.

\subsection{Momentum current density of NRQED}

After introducing the NRQED in the Coulomb gauge, one now discuss its momentum current density. Clearly, the EMT of NRQED must include the non-relativistic reduction of the QED EMT, which we call the ``tree-level'' EMT. In addition, the naive EMT in theories with non-trivial UV structure can receive quantum corrections, which must be included as counter terms in the effective operator. For example, in the scalar $\phi^4$ theory in 4D, in order for off-shell matrix elements of EMT to be finite, one must add local counter-term of the form $ d(\partial^{\mu}\partial^{\nu}-g^{\mu\nu}\partial^2) \phi^2$ to the naive EMT, where $d$ is divergent order-by-order in perturbation theory. In our case,
we are only interested in corrections at one-loop that are spin-independent.
As we will show in Sec.III that to order $\frac{\alpha}{m_e}$, the EMT of NRQED has the form
\begin{align}\label{eq:Tmatch}
    T^{ij}_{\rm NRQED}=T^{ij}_{\rm tree}+d_0(\partial^i\partial^j-\delta^{ij}\partial^2)\Psi^{\dagger}\Psi  + {\cal O}(\alpha^2)\ ,
\end{align}
where $d_0$ depends on the UV regulator for the NRQED, but is IR insensitive.

We now consider all contributions to $T^{ij}_{\rm tree}$ and their Feynman rules. By the standard NR reduction on the energy-momentum tensor of QED, one can obtain the ``classical'' EMT for NRQED,
\begin{align}\label{eq:Tclass}
T^{ij}_{\rm tree}=T^{ij}_{e}+T_{\gamma}^{ij}+T_{\gamma p}^{ij}+T_{p}^{ij}\ .
\end{align}
where
\begin{enumerate}
    \item $T^{ij}_{e}$ is the electron part of the EMT. For the non-relativistic particle, by performing the non-relativistic reduction on the full QED,  one obtains:
\begin{align}
    &T_{e}^{ij}=-\frac{1}{4m_e}\Psi^{\dagger}D^{i}D^j\Psi-\frac{1}{4m_e}(D^iD^j\Psi)^{\dagger}\Psi\nonumber \\
    &+\frac{1}{4m_e}(D^i\Psi)^{\dagger}(D^j\Psi)+\frac{1}{4m_e}(D^j\Psi)^{\dagger}(D^i\Psi) \ .
\end{align}
Expanding all the covariant derivatives, it can be further decomposed as
\begin{align}\label{eq:Tphi}
T^{ij}_{e}=T^{ij}_{e0}+T^{ij}_{e1}+T^{ij}_{e2} \ ,
\end{align}
where
\begin{align}
&T^{ij}_{e0}=-\frac{1}{4m_e}\Psi^{\dagger}\partial^{i}\partial^j\Psi-\frac{1}{4m_e}\partial^{i}\partial^j\Psi^{\dagger}\Psi\nonumber \\& +\frac{1}{4m_e}\partial^{i}\Psi^{\dagger}\partial^{j}\Psi+\frac{1}{4m_e}\partial^j\Psi^{\dagger}\partial^i\Psi \nonumber \\
    &T^{ij}_{e1}=\frac{ie}{2m_e}A^i(\Psi^{\dagger}\partial^j\Psi-\partial^j\Psi^{\dagger}\Psi)-(i\leftrightarrow j)\nonumber \\
    &T^{ij}_{e2}=\frac{e^2}{m_e}A^iA^j\Psi^{\dagger}\Psi \ .
\end{align}
Notice the appearance of $T^{ij}_{e2}$, which is absent in the full QED.
\item $T^{ij}_{\gamma}$ is the standard contribution from the photon-field $A^{\mu}=(A^0,\vec{A})$,
  \begin{align}\label{eq:Tphoton}
   T_{\rm \gamma}^{\mu\nu}=-F^{\mu\rho}F^{\nu}_{\rho}+\frac{g^{\mu\nu}}{4}F^2 \ .
  \end{align}
It is convenient to decompose the $T^{ij}_{\gamma}$ into pure-Coulomb, mixed and pure radiative parts
\begin{align}
    T^{ij}_{\gamma}=T^{ij}_{\gamma\parallel}+T^{ij}_{\gamma\parallel \perp}+T^{ij}_{\gamma\perp} \ ,
\end{align}
with
\begin{align}
    &T^{ij}_{\gamma\parallel}=-\partial^iV_e\partial^jV_e+\frac{1}{2}\delta^{ij}\partial^kV_e\partial^k V_e  \ ,  \label{eq:Tcou}\\
    &T^{ij}_{\gamma\parallel\perp}=2\partial^{(i}V_e\partial^0A^{j)}-\delta^{ij}\partial^kV_e\partial^0A^k \ .\label{eq:Tmixed}
\end{align}
where $A^{(ij)}=\frac{1}{2}\left(A^{ij}+A^{ji}\right)$ denotes the standard symmetrization. For the pure radiative part $T^{ij}_{\gamma \perp}$, one can further decompose it into electric and magnetic part
\begin{align}\label{eq:Tradia}
    T^{ij}_{\gamma\perp}=T^{ij}_{\gamma E}+T^{ij}_{\gamma B} \ ,
\end{align}
with
\begin{align}
    T^{ij}_{\gamma E}=&-\partial^0A^i\partial^0A^j+\frac{\delta^{ij}}{2}\partial^0A^k\partial^0 A^k \ , \label{eq:TE} \\
    T^{ij}_{\gamma B}=&(\partial^iA^k-\partial^kA^i)(\partial^jA^k-\partial^kA^j) \nonumber \\
    &-\frac{\delta^{ij}}{2}(\partial^kA^l\partial^kA^l-\partial^kA^l\partial^lA^k) \ . \label{eq:TB}
\end{align}
This decomposition will be used later.
\item The $T^{ij}_{\gamma p}$ is the mixed contribution between the photon field and the proton's electric field. More precisely, it can be further decomposed as
\begin{align}
T^{ij}_{\gamma p}=T^{ij}_{\perp p}+T^{ij}_{\parallel p} \ ,
\end{align}
where $T^{ij}_{\parallel p}$ is the mixing between the electron's Coulomb field and the proton's electric field
\begin{align}
    T_{\parallel p}^{ij}=\delta^{ij}\nabla V_p \cdot \nabla V_e- \partial^i V_e \partial^j V_p-\partial^iV_p\partial^jV_e \ ,
\end{align}
while $T^{ij}_{\perp p}$ is the mixing between radiative field and the proton's electric field
\begin{align}\label{eq:mixinter}
    T^{ij}_{\perp p}=\partial^iV_p\partial^0A^j+\partial^jV_p\partial^0A^i-\delta^{ij}\partial^kV_p\partial^0A^k \ .
\end{align}
\item Finally, $T^{ij}_p$ is the energy momentum tensor of the proton, which can be calculated using its classical electric field $E_p^i=\frac{er^i}{4\pi r^3}$ as
\begin{align}
T^{ij}_{p}(\vec{r})=(\delta^{ij}\partial^2-\partial^i\partial^j)\frac{\alpha}{32\pi r^2}\ ,
\end{align}
which translate to momentum space as
\begin{align}\label{eq:Tp}
T^{ij}_{p}(\vec{q})=(q^iq^j-\delta^{ij}q^2)\frac{\alpha\pi}{16|q|} \ .
\end{align}
It is transverse by itself.
\end{enumerate}
One can show that sandwiched between static states with equal energies, the above $T_{\rm tree }^{ij}$ is conserved.  In fact, in Appendix A we will show that $T^{ij}_e+T^{ij}_{\gamma}+T^{ij}_{\parallel p}$ and $T^{ij}_{\perp p}$ are conserved separately for the spherical symmetric ground state where $V_e(p)=V_e(|p|)$ .

\begin{figure}[htb]	
		\begin{subfigure}{.17\textwidth}
		\includegraphics[width=\linewidth]{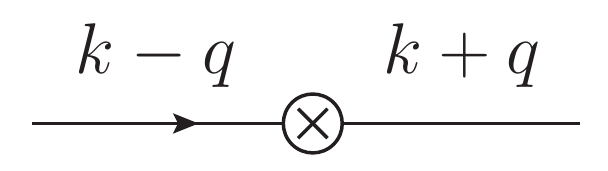}
		\caption{}
		\label{fig:Ferminkine}
	\end{subfigure}
   \begin{subfigure}{.18\textwidth}
		\includegraphics[width=\linewidth]{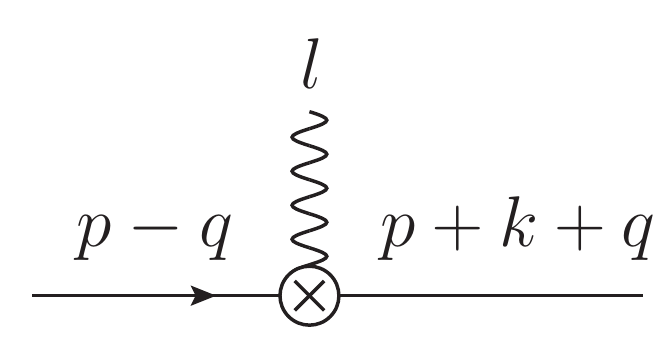}
		\caption{}
		\label{fig:Fermionphoton}
	\end{subfigure}
		\begin{subfigure}{.18\textwidth}
	\centering
	\includegraphics[width=\linewidth]{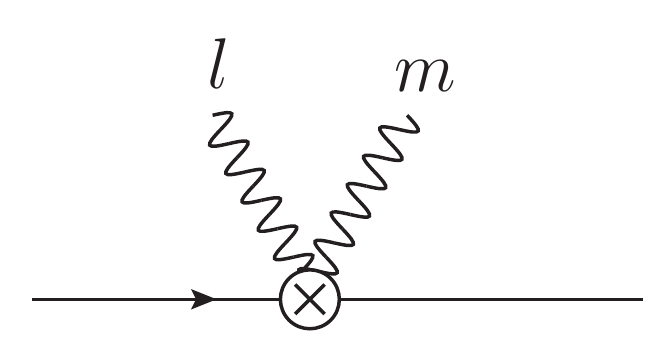}
	\caption{}
	\label{fig:Fermiontadpole}
\end{subfigure}	
\caption{Vertices corresponding to different terms for $T^{ij}_{e}$ in Eq.~(\ref{eq:Tphi}). Fig.~\ref{fig:Ferminkine} corresponds to $T^{ij}_{e0}$, Fig.~\ref{fig:Fermionphoton} corresponds to$T^{ij}_{e1}$ and Fig.~\ref{fig:Fermiontadpole} corresponds to  $T^{ij}_{e2}$.}
		\label{fig:fermionrules}
\end{figure}

\begin{figure}[htb]	
		\begin{subfigure}{.2\textwidth}
		\centering
		\includegraphics[width=\linewidth]{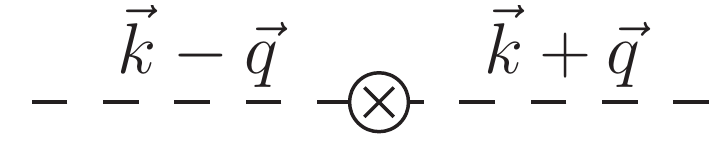}
		\caption{}
		\label{fig:pureCoulomb}
	\end{subfigure}

  \begin{subfigure}{.24\textwidth}
		\centering
		\includegraphics[width=\linewidth]{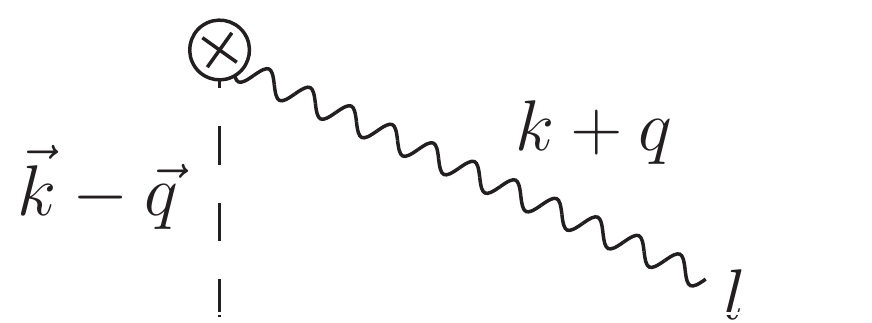}
		\caption{}
		\label{fig:mixedvertex}
	\end{subfigure}
		\begin{subfigure}{.2\textwidth}
	\centering
	\includegraphics[width=\linewidth]{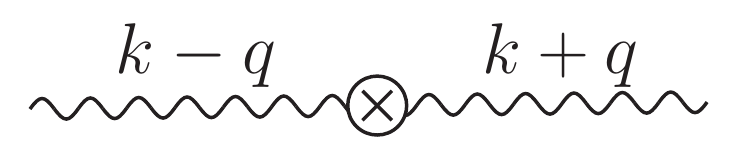}
	\caption{}
	\label{fig:pureradiative}
\end{subfigure}	
\caption{Vertices corresponding to different terms for $T^{ij}_{\gamma}$ in Eq.~(\ref{eq:Tphoton}). Fig.~\ref{fig:pureCoulomb} corresponds to $T^{ij}_{\gamma \parallel}$, Fig.~\ref{fig:mixedvertex} corresponds to $T^{ij}_{\gamma \parallel \perp}$ and Fig.~\ref{fig:pureradiative} corresponds to $T^{ij}_{\gamma \perp}$. }
		\label{fig:photonrules}
\end{figure}

 For convenience of the reader, we collect all the vertices for the above momentum current density in Fig.~\ref{fig:fermionrules} and Fig.~\ref{fig:photonrules}. More precisely, for the fermion part by simply taking the matrix element of the various terms in Eq.~(\ref{eq:Tphi}) in plan wave states one has
\begin{align}
    &\text{\rm Fig.~\ref{fig:Ferminkine}}=k^ik^j \ , \\
    &\text{\rm Fig.~\ref{fig:Fermionphoton}}=-\frac{e}{2m_e}\delta^{il}(2p+k)^j-\frac{e}{2m_e}\delta^{jl}(2p+k)^i \ , \\
    &\text{\rm Fig.~\ref{fig:Fermiontadpole}}=\frac{e^2}{m_e}(\delta^{il}\delta^{jm}+\delta^{jl}\delta^{im}) \ .
\end{align}
Similarly, for the photon part one has
\begin{align}
    &\text{\rm Fig.~\ref{fig:pureCoulomb}}=-2(k-q)^{(i}(k+q)^{j)}+\delta^{ij}(\vec{k}^2-\vec{q}^2) \ , \\
    &\text{\rm Fig.~\ref{fig:mixedvertex}}=2k^0(k-q)^{(i}\delta^{j)l}-\delta^{ij}k^0(k-q)^l \ ,
\end{align}
Using Eq.~(\ref{eq:TE}) and Eq.~(\ref{eq:TB}), Fig.~\ref{fig:pureradiative} can simply be obtained by taking matrix elements of $T^{ij}_{E}+T^{ij}_{B}$ in free photon states. Since the resulting expression is quite long and since in the calculation one only needs $T^{ii}_{\gamma \perp}$ and $q^iT^{ij}_{\gamma \perp}q^j$, which simplifies considerably, we will not provide the explicit formulas here.

\subsection{Leading-order momentum-current form factor}

\begin{figure}[htb]	
		\begin{subfigure}{.25\textwidth}
		\centering
		\includegraphics[width=\linewidth]{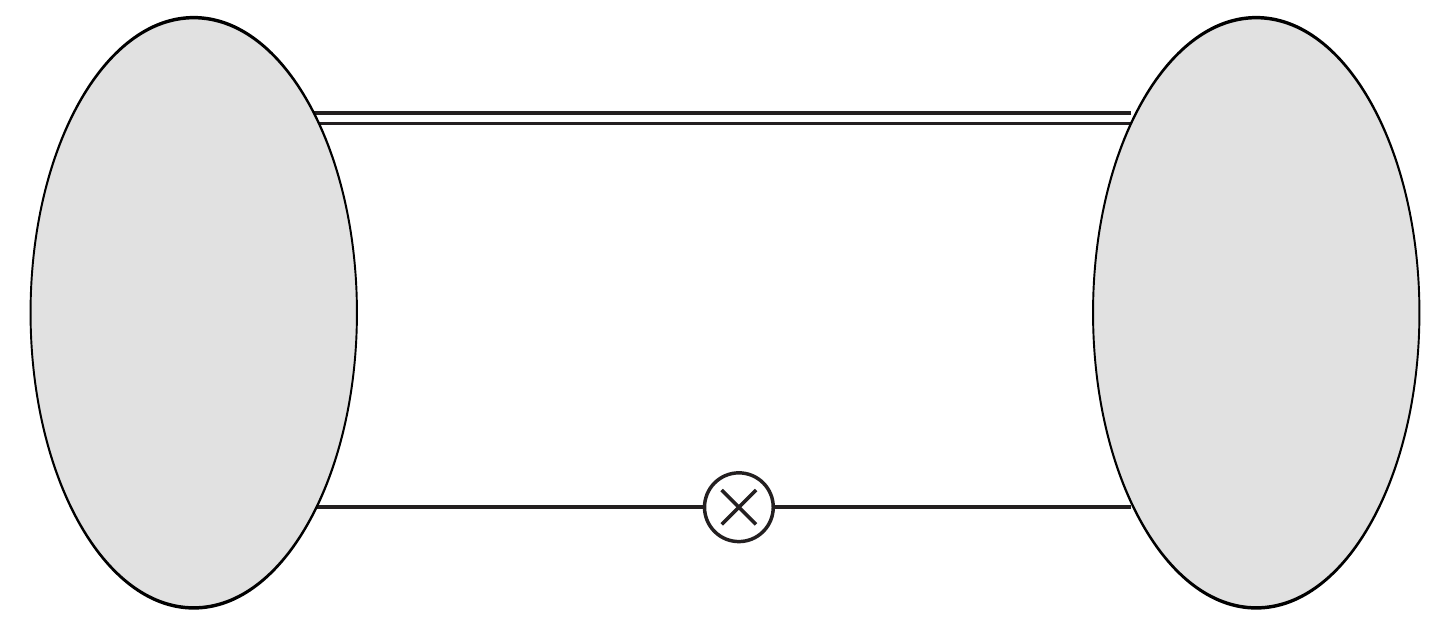}
		\caption{}
		\label{fig:electron1}
	\end{subfigure}

  \begin{subfigure}{.25\textwidth}
		\centering
		\includegraphics[width=\linewidth]{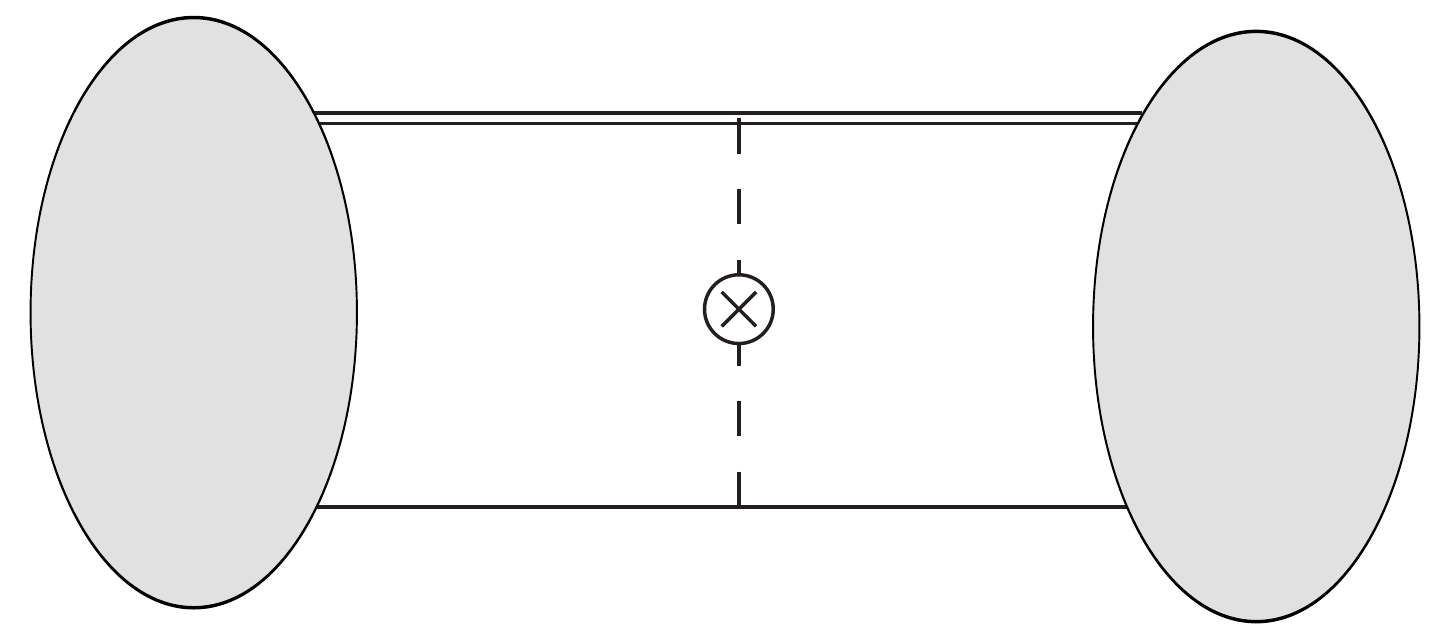}
		\caption{}
		\label{fig:inter1}
	\end{subfigure}
		\begin{subfigure}{.25\textwidth}
	\centering
	\includegraphics[width=\linewidth]{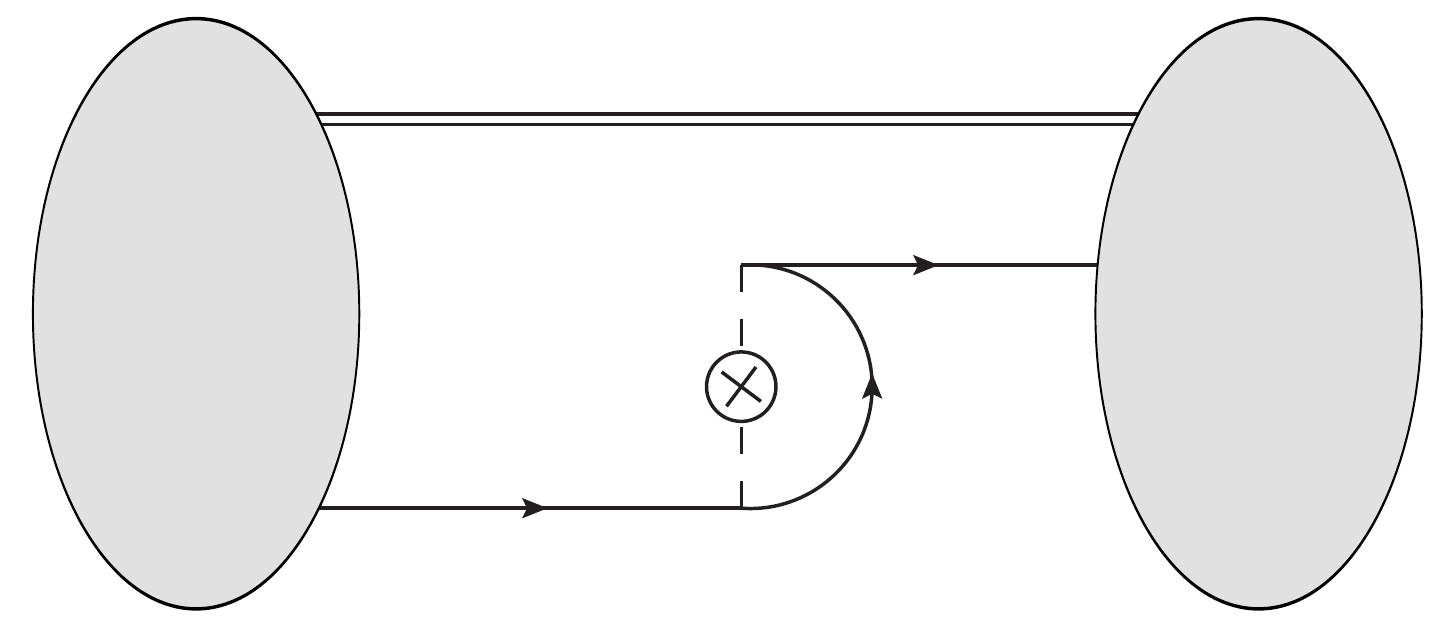}
	\caption{}
	\label{fig:Coulomself1}
\end{subfigure}	
		\begin{subfigure}{.25\textwidth}
	\centering
	\includegraphics[width=\linewidth]{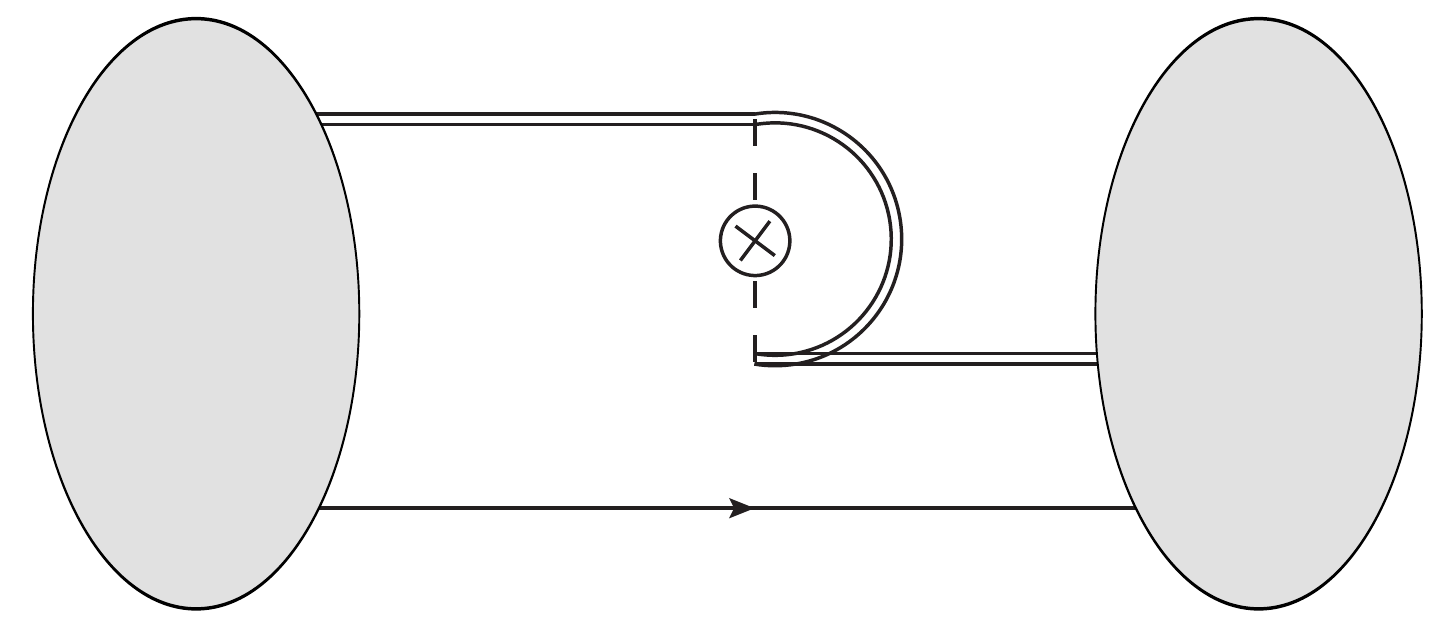}
	\caption{}
	\label{fig:Coulombpro}
\end{subfigure}
\caption{The order-${\cal O}(1)$ contributions to $T^{ij}$ for a bound state.  Dashed lines represent Coulomb photons and crossed circles denote the operator insertions. Notice the infrared divergences for $C(q)$ at $q=0$ are cancelled between the interference and single electron and proton contributions.}
		\label{fig:inter}
\end{figure}

Given the above momentum current operator, one can calculate its form factor in the ground state of hydrogen atom.  The contributions to the leading order are shown in Fig.~\ref{fig:inter}. More explicitly, in Fig.~\ref{fig:electron1} one has the electron kinetic contribution, in Fig.~\ref{fig:inter1} one has the interference contribution between Coulomb photons emitted from the electron and the proton. These two terms are conserved when added together. Finally, in Fig.~\ref{fig:Coulomself1} one has the Coulomb-photon self-energy contributions from the electron and proton, respectively. Since there is no UV divergence at this order, all calculations can be performed in $D=3$ directly.

More explicitly, the fermionic contribution in Fig.~\ref{fig:electron1} can be shown as
\begin{align}\label{eq:kine}
 \langle T^{ij}_{e0}(\vec{x}) \rangle =-\frac{1}{4m_e}\left(\psi^{\dagger}_0\partial^{i}\partial^{j}\psi_0-\partial^{i}\psi^{\dagger}_0\partial^{j}\psi_0+{\rm c.c}\right) \ .
\end{align}
where $\langle T^{ij}\rangle \equiv \langle 0|T^{ij}|0 \rangle$ denotes the matrix element in the ground state $|0\rangle$ of the hydrogen atom with wave function $\psi_0$.
Furthermore, interference contribution, Fig.~\ref{fig:inter1} be calculated as
\begin{align}
    \langle T^{ij}_{||p}(\vec{x})\rangle=\delta^{ij}\nabla V_p \cdot \nabla V_e- \partial^i V_e \partial^j V_p-\partial^iV_p\partial^jV_e \ ,
\end{align}
where the static potential $V_e$ induced by the electron reads
\begin{align}
\nabla^2 V_e(\vec{x})=e|\psi_0(\vec{x})|^2 \ ,
\end{align}
 One can show that the quantum mechanical contribution
 $\langle T^{ij}\rangle_{\rm e0}+\langle T^{ij}\rangle_{\rm ||p}$ is conserved by itself.

 After Fourier-transformation to the momentum space,
 \begin{align}\label{eq:CQM}
    &\langle T^{ij}\rangle_{ e0+\parallel p}(\vec{q})=(q^iq^j-\delta^{ij}q^2)\frac{C_{ e0+\parallel p}(q)}{m_e} \ ,\\
    &\frac{C_{e0+\parallel p}(q)}{m_e}=\frac{1}{2m_e(\frac{q^2}{\alpha^2m_e^2}+4)}-\frac{\alpha}{4|q|}\left(\frac{\pi}{2}-{\rm Arctan} \frac{q}{2\alpha m_e}\right)  \ .
  \end{align}
The resulting $C_{\rm QM}(q)$ contains a a Coulomb tail $\frac{\pi \alpha}{8|q|}$.
In order to cancel it, one must add the electron and proton Coulomb self-energy contributions, which we will show to be conserved by itself.

For the bound state, the Coulomb contribution in Fig.~\ref{fig:Coulomself1} can be calculated using the Feynman rules provided above . It turns out that the self-energy bubble for the Coulomb insertion is independent of the incoming/out going electron/proton momentum,  and contributes exactly as the free-electron, which will be further dressed by the momentum dependency of the bound state wave function for the external electron and proton.  The form factor from the electron therefore reads
\begin{align}
   \frac{ C_{\gamma \parallel } (q)}{m_e}= \frac{\alpha \pi}{16|q|}\times \frac{16\alpha^4}{(\frac{q^2}{m_e^2}+4\alpha^2)^2} \ ,
\end{align}
where the first factor $\frac{\alpha \pi}{16|q|}$ is just the free-electron contribution, and the second factor is nothing but the dressing in the bound-state wave function
\begin{align}
\frac{16\alpha^4}{(\frac{q^2}{m_e^2}+4\alpha^2)^2}=\int \frac{d^3\vec{p}}{(2\pi)^3} \psi^{\dagger}_0(\vec{p}-\frac{\vec{q}}{2})\psi_0(\vec{p}+\frac{\vec{q}}{2}) \ ,
\end{align}
where $p\pm \frac{q}{2}$ are the external momentum of the self-energy bubble.  Finally, the contribution of the proton can be obtained from Eq.~(\ref{eq:Tp}) as
\begin{align}
    \frac{C_{\rm p} (p)}{m_e} = \frac{\alpha \pi}{16|q|} \ .
\end{align}
Equivalently, this can also be calculated from the $T^{ij}_{\rm Coulomb}$ for an infinitely heavy source field, which explains the representation in Fig.~\ref{fig:Coulombpro}.

In conclusion, in the region $|q|\le{\cal O} (\alpha m_e)$, the $C$ form factor of the hydrogen atom reads
\begin{align}\label{eq:hydrogenfinal}
  C_H(q)=C_{e0}(q)+C_{\parallel p}(q)+C_e(q)+C_p(q) \ .
\end{align}
From these, the tensor monopole moment  for the hydrogen atom is
\begin{align}
    \tau_{\rm H}=\frac{C_{\rm H}(0)}{m_e}\equiv g_H\tau_0=\tau_0[1 +{\cal O}(\alpha \ln \alpha)]\ .
\end{align}
where $\tau_0 = \hbar^2/4m_e$. Therefore, $g_H=1$, except for a small correction of order $\alpha$, a result with opposite sign from a point-like boson.

\subsection{Diagrams and power-counting in ${\cal O}(\alpha)$}

\begin{figure}[htb]	
		\begin{subfigure}{.4\textwidth}
		\centering
		\includegraphics[width=\linewidth]{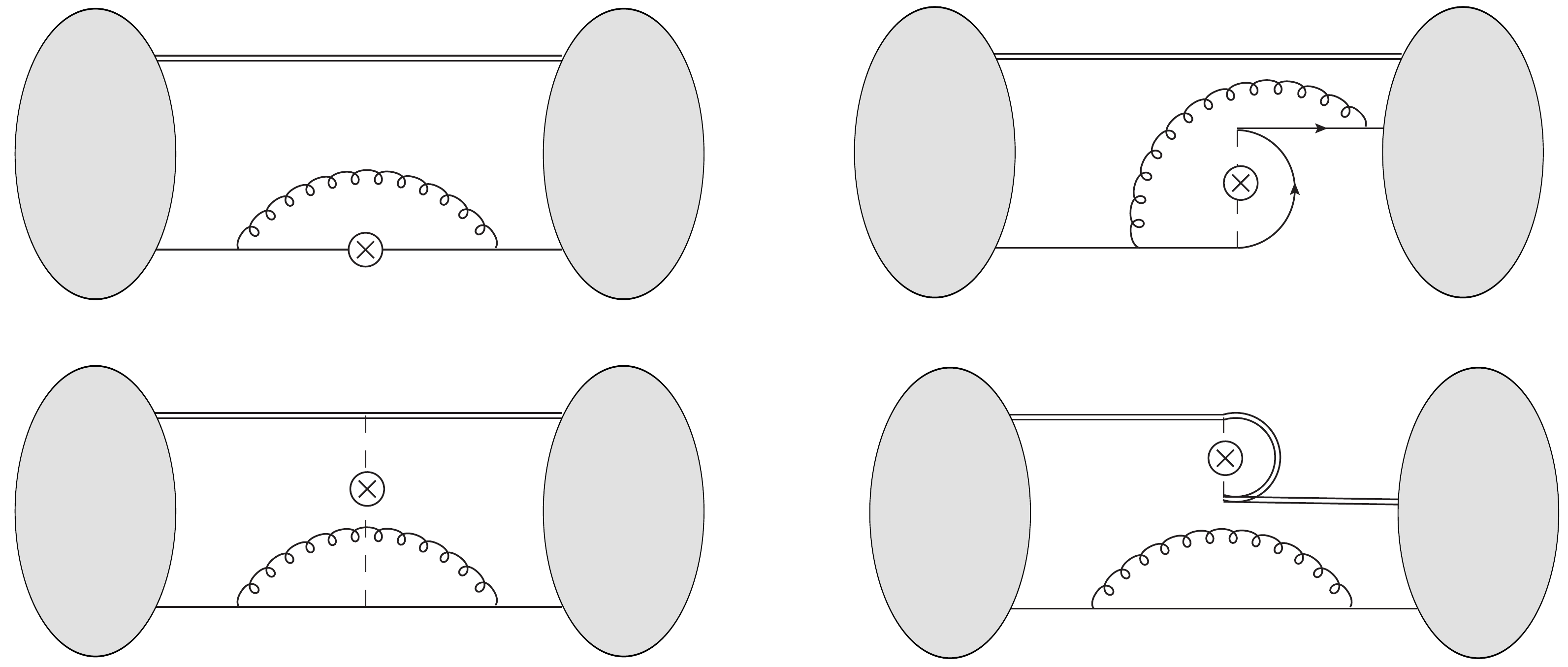}
		\caption{}
		\label{fig:fermionver}
	\end{subfigure}
\begin{subfigure}{.4\textwidth}
		\centering
		\includegraphics[width=\linewidth]{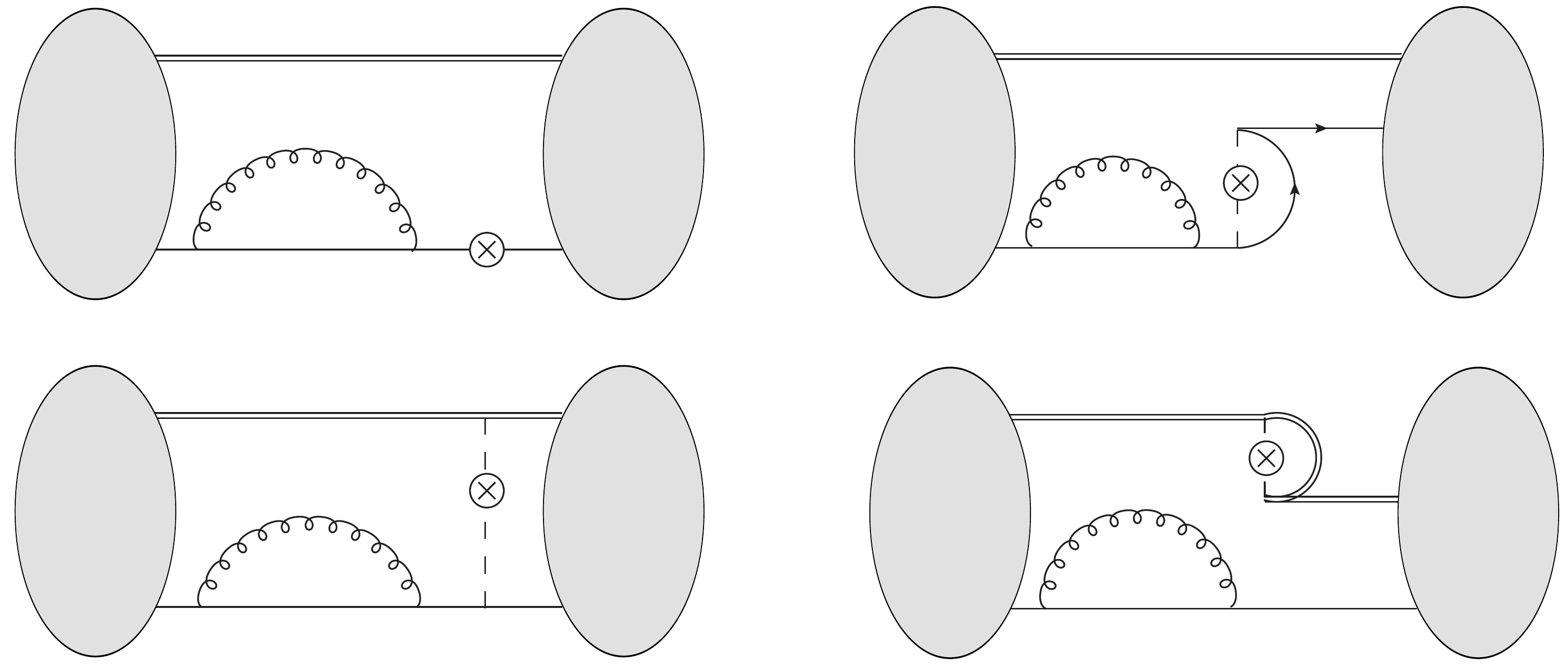}
		\caption{}
		\label{fig:fermionver1}
	\end{subfigure}
\begin{subfigure}{.2\textwidth}
		\centering
		\includegraphics[width=\linewidth]{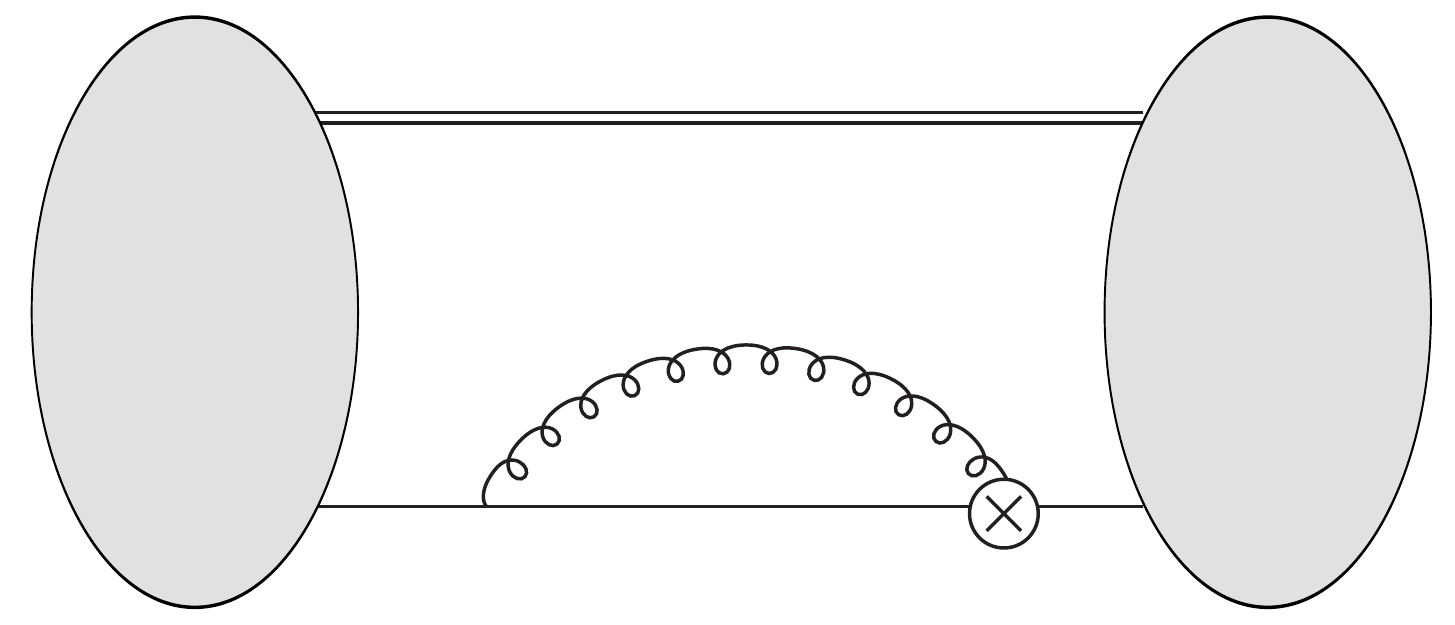}
		\caption{}
		\label{fig:fermionop}
	\end{subfigure}
\begin{subfigure}{.2\textwidth}
		\centering
		\includegraphics[width=\linewidth]{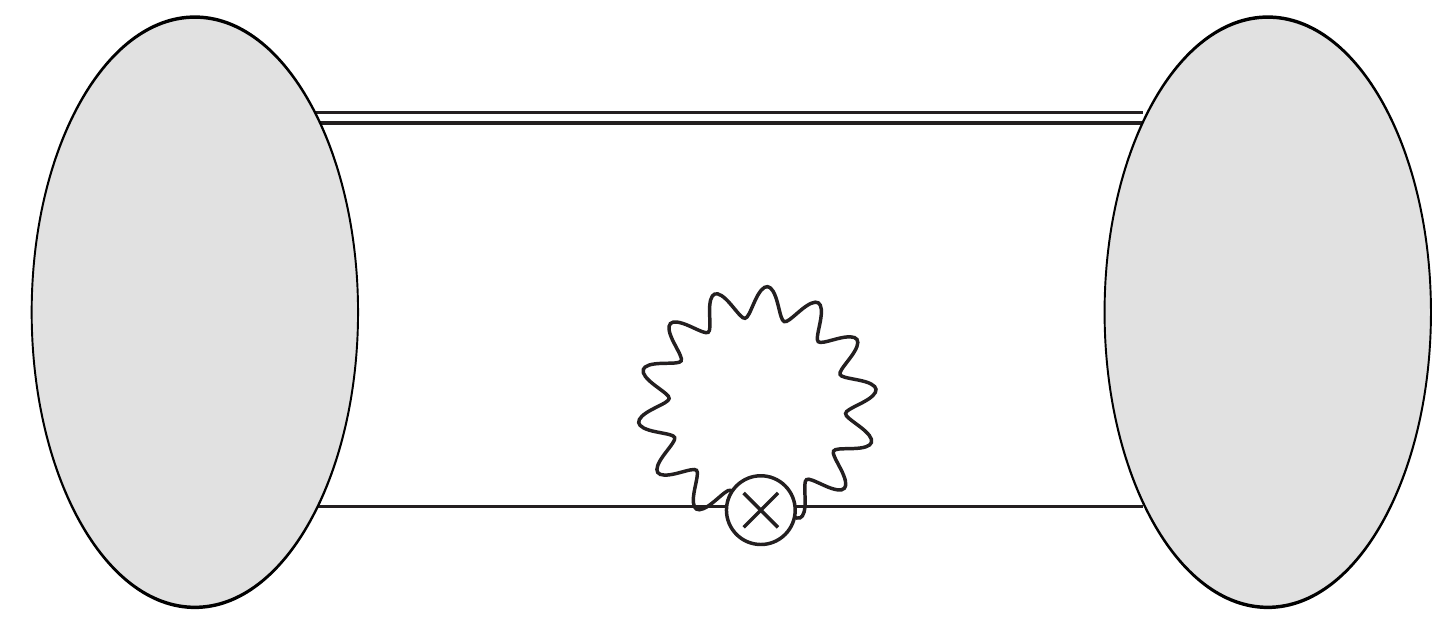}
		\caption{}
		\label{fig:fermiontadlamb}
	\end{subfigure}

\caption{The one-loop fermionic contributions to $T^{ij}$ for a bound state, shown in old-fashioned perturbation theory. Dashed lines represent Coulomb photons and crossed circles denote the operator insertions. Notice that Fig.~\ref{fig:fermionver1} contains two types of contributions. All of them are sub-leading in $\alpha$ and will not be calculated. }
		\label{fig:NLOf}
\end{figure}

\begin{figure}[htb]	
		\begin{subfigure}{.25\textwidth}
		\centering
		\includegraphics[width=\linewidth]{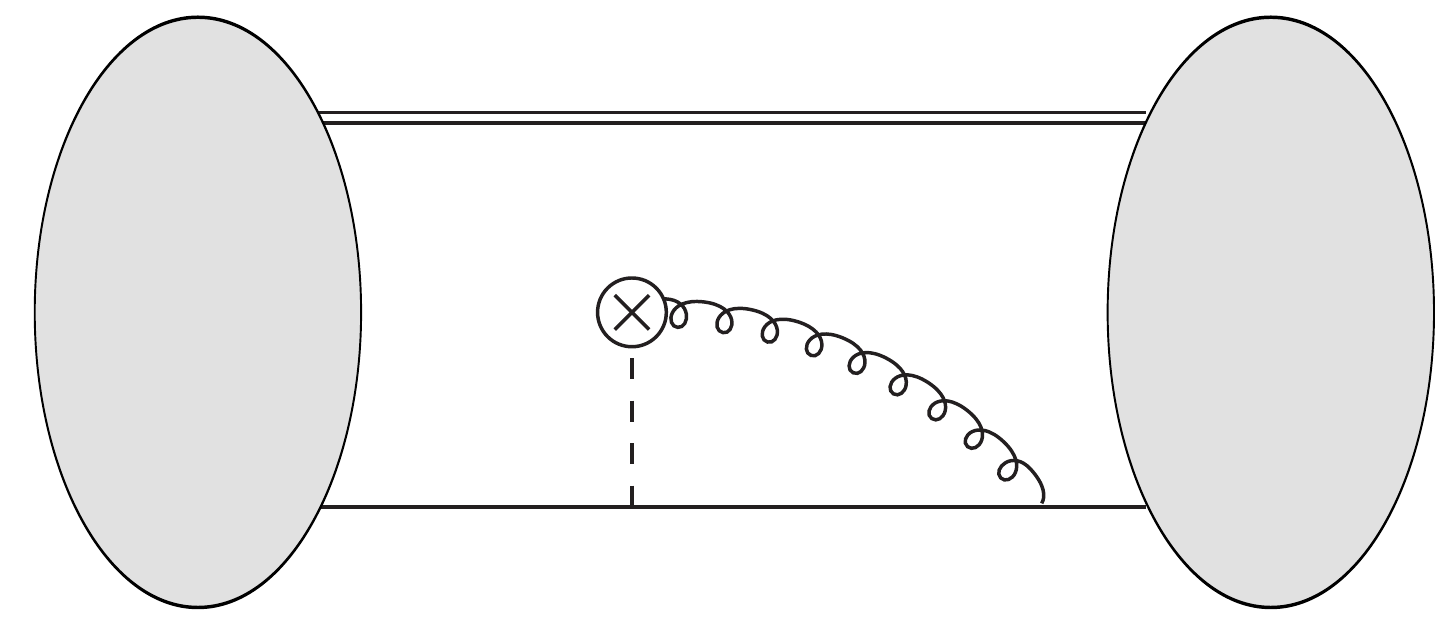}
		\caption{}
		\label{fig:lambmix}
	\end{subfigure}
\begin{subfigure}{.25\textwidth}
		\centering
		\includegraphics[width=\linewidth]{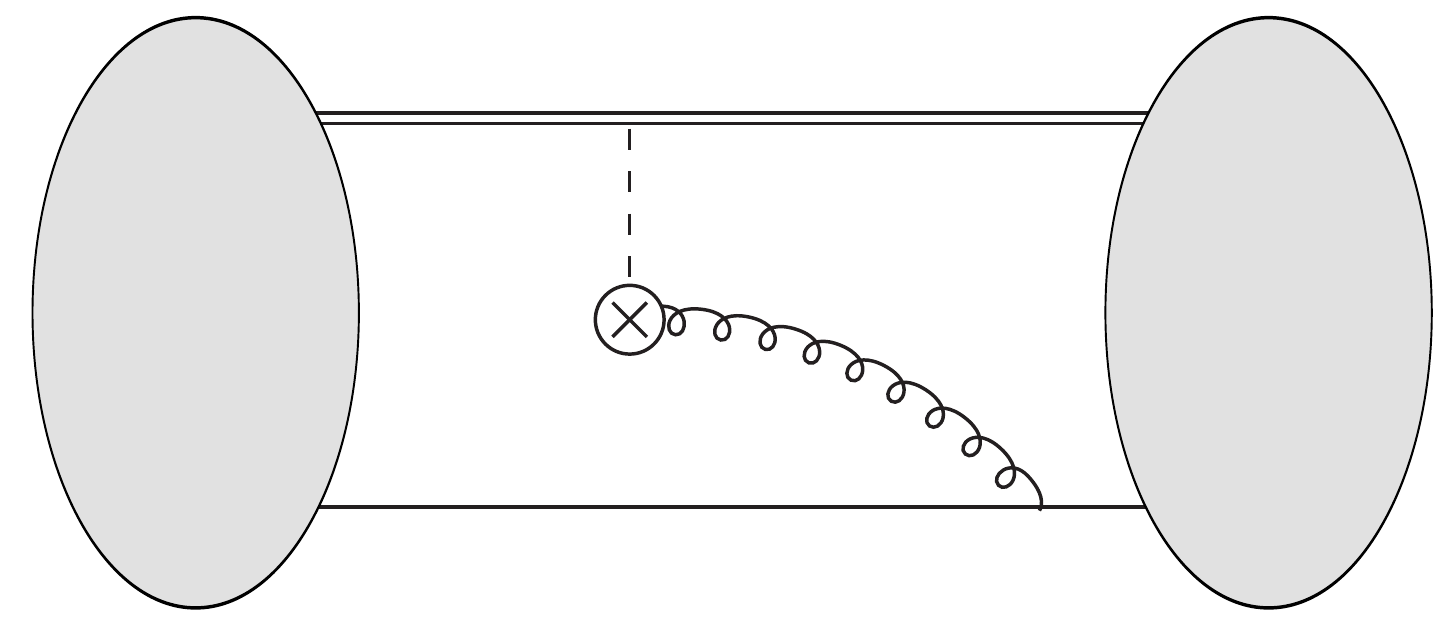}
		\caption{}
		\label{fig:lambmixinter}
	\end{subfigure}
 \begin{subfigure}{.25\textwidth}
	\centering
	\includegraphics[width=\linewidth]{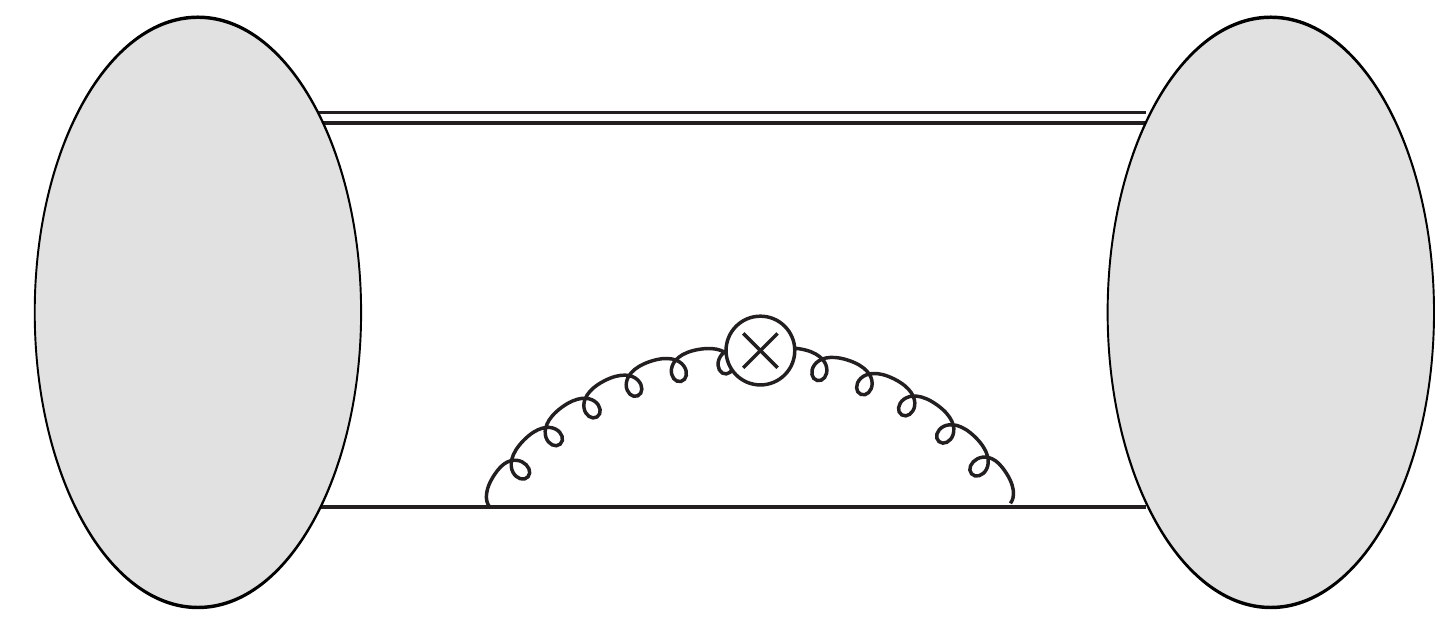}
	\caption{}
	\label{fig:lambradia}
\end{subfigure}	
\begin{subfigure}{.25\textwidth}
	\centering
	\includegraphics[width=\linewidth]{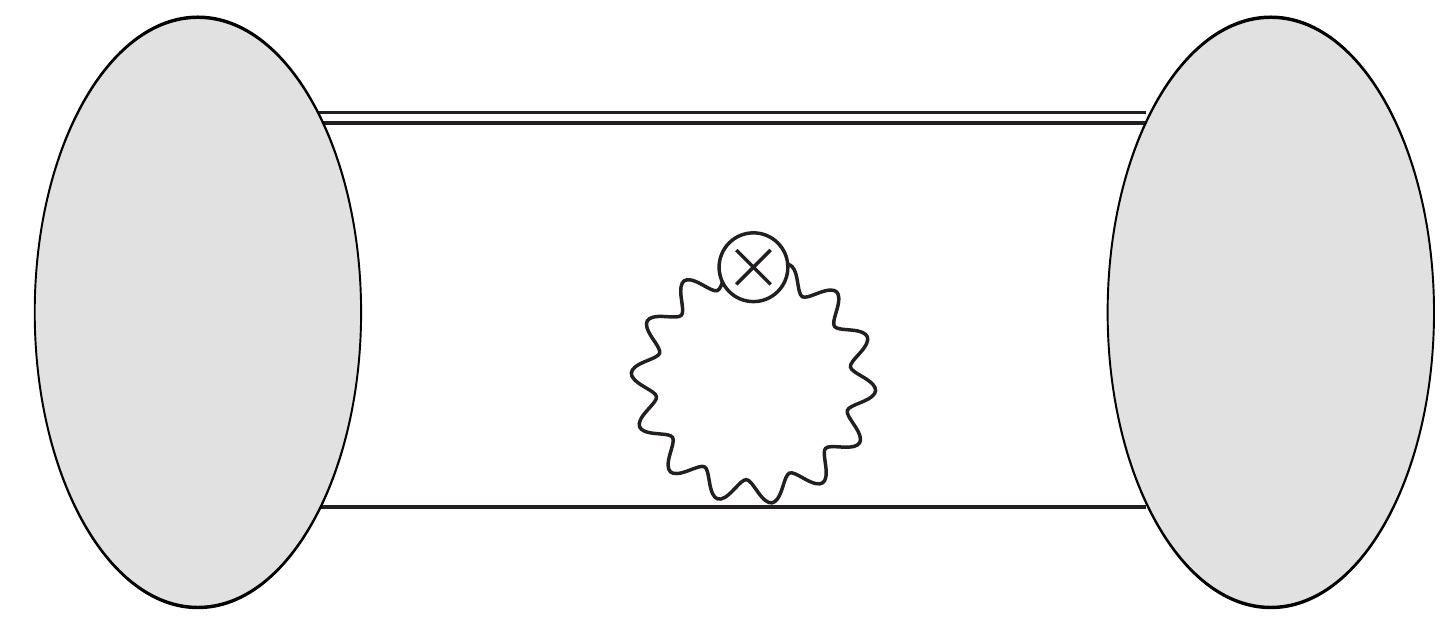}
	\caption{}
	\label{fig:lambtad}
\end{subfigure}	

\caption{The order-${\cal O}(\alpha)$ contributions to $T^{ij}_{\gamma}+T^{ij}_{\gamma p}$ for a bound state. Dashed lines represent Coulomb photons and crossed circles denote the operator insertions.}
		\label{fig:NLObound}
\end{figure}

To next-to-leading order in radiative corrections, relevant diagrams for the form factor
are shown in Fig.~\ref{fig:NLOf} and Fig.~\ref{fig:NLObound}. There are ten fermionic diagrams and four photonic one. We show that only the photonic contributions in Fig.~\ref{fig:NLObound} are relevant
in the soft and ultra-soft regions and contributes to $\tau_H$ to order $\alpha$.

For this purpose, it is convenient to follow the previous subsection to
introduce the leading-order momentum current independent of the external states,
\begin{align}\label{eq:Tleading}
&T_0^{ij}(\vec{q})=T^{ij}_{e0}(\vec{q})+T_{\parallel p}^{ij}(\vec{q})\nonumber \\
&+(q^iq^j-\delta^{ij}q^2)\frac{\alpha\pi}{16|q|}\Psi^{\dagger}\Psi(q)+(q^iq^j-\delta^{ij}q^2)\frac{\alpha\pi}{16|q|}   \ .
\end{align}
It is easy to show by using the standard power-counting rules Eq.~(\ref{eq:powerphi}) and Eq.~(\ref{eq:powercs}) that for non-relativistic wave functions for electrons, that one has the basic power-counting rule
\begin{align}
\langle M'|T^{ij}_{\rm 0}(\vec{q})|M\rangle={\cal O}(\alpha^2)+ {\cal O}(\alpha)q+{\cal O}(1)q^2 \ ,
\end{align}
where $q$, $q^2$ denotes the expansion of $\langle M'|T_{\rm 0}^{ij}(\vec{q})|M\rangle$ to linear and quadratic orders in $q$, respectively.

Given the above, it is not hard to show that all the fermionic diagrams are sub-leading in $\alpha$. For example, for the four diagrams shown in Fig.\ref{fig:fermionver}, the Feynman rule for old-fashioned perturbation theory shows that they combine to produce
\begin{align}
&\langle T^{ij}_{0}\rangle_{\rm 6a} (\vec{q})=e^2\mu^{2\epsilon}\sum_{M,M'}\int \frac{d^D\vec{k}}{(2\pi)^D}\nonumber \\
&\frac{(1-\frac{1}{D})\vec{v}_{0M'}(\vec{k})\cdot \vec{v}_{M0}(\vec{k})}{(E_0-|\vec{k}|-E_M)(E_0-|\vec{k}|-E_{M'})2|\vec{k}|}\langle M'|T_{0}^{ij}(\vec{q})|M\rangle \ ,
\end{align}
where the $T_{\rm 0}^{ij}(\vec{q})$ is defined in Eq.~(\ref{eq:Tleading}), including the proton part when $M=M'$.  It is easy to see that for $D=3$, when $|\vec{k}|={\cal O}(\alpha m_e)$ or $|\vec{k}|={\cal O}(\alpha^2 m_e)$, the two energy denominators and one phase-space measure $2|\vec{k}|$ for the photon contributes to $(\alpha m_e)^{-3}$ or $(\alpha m_e)^{-6}$, which is always canceled by the integration measure $\int d^3 \vec{k}={\cal }(\alpha m_e)^3$ or $(\alpha m_e)^6$, respectively. The two form-factors for the velocity operators will contributes to $\alpha^2$, while the matrix element $\langle M'|T_{\rm 0}^{ij}(q)|M\rangle $ as shown above will contributes to ${\cal O}(\alpha^2)$ at order ${\cal O}(q^0)$ and ${\cal O}(1)$ to ${\cal O}(q^2)$. Therefore, together with the overall $e^2$, Fig.~\ref{fig:fermionver} will contributes at order ${\cal O}(\alpha^3)$ to the coefficients of $q^2$, and to  ${\cal O}(\alpha^5)$ for coefficients of $q^0$, therefore not relevant for our calculation. Similar argument can be used to show that diagrams in Fig.~\ref{fig:fermionver1} and Fig.~\ref{fig:fermionop} ($T^{ij}_{e1}$) will be irrelevant to NLO as well. The last diagram Fig.~\ref{fig:fermiontadlamb}, due to the $\frac{e^2}{m_e}\Psi^{\dagger}\Psi A^iA^j$ term in $T^{ij}_{e}$, reads
\begin{align}
&\langle T^{ij}_{e2}\rangle_{6c}(q)=\int \frac{d^Dp}{(2\pi)^D} \psi^{\dagger}_0(p-\frac{q}{2})\psi_0(p+\frac{q}{2})\nonumber \\
&\times \frac{e^2}{2m_e}\mu^{2\epsilon}\int \frac{d^D\vec{k}}{(2\pi)^D|\vec{k}|}\bigg(\delta^{ij}-\frac{k^ik^j}{\vec{k}^2}\bigg) \ ,
\end{align}
which is proportional to a $q$-independent dimensionless integral and vanishes identically in dimensional regularization.

More generally, the power-counting for arbitrary diagram with interaction vertices from the Lagrangian in Eq.~(\ref{eq:NRLag}) can be performed as follows. We first consider the one-body irreducible (1PI) diagrams containing the operator insertion, but without insertion of self-energy type bubbles on external electron legs. Then it is easy to see that for arbitrary 1PI diagram one has
\begin{align}
    \langle T_{0}^{ij}\rangle_{\rm s}=\bigg(q^2{\cal O}(1)+q{\cal O}(\alpha)+{\cal O}(\alpha^2)\bigg)\alpha^{\frac{3}{2}n_{V3}+2n_{V_4}} \ , \\
    \langle T_{0}^{ij}\rangle_{\rm us}=\bigg(q^2{\cal O}(1)+q{\cal O}(\alpha)+{\cal O}(\alpha^2)\bigg)\alpha^{\frac{3}{2}n_{V3}+3n_{V_4}} \ ,
\end{align}
where $n_{V3}$ and $n_{V_4}$ denotes the numbers of electron-photon triple and seagull vertices, respectively, and `s', `us' denotes that all the photons in the diagram are soft or ultra-soft.
For the fermion-photon mixed operator, one has
\begin{align}
\langle T_{e1}^{ij}\rangle_{\rm s}=\bigg(q^2{\cal O}(1)+q{\cal O}(\alpha)+{\cal O}(\alpha^2)\bigg) \alpha^{\frac{3}{2}n_{V3}+2n_{V_4}+\frac{1}{2}} \ , \\
    \langle T_{e1}^{ij}\rangle_{\rm us}=\bigg(q^2{\cal O}(1)+q{\cal O}(\alpha)+{\cal O}(\alpha^2)\bigg)\alpha^{\frac{3}{2}n_{V3}+3n_{V_4}+\frac{3}{2}} \ .
\end{align}
The leading contribution is shown in Fig. 6c. For the fermion-tadpole operator,
\begin{align}
\langle T_{\rm e2}^{ij}\rangle_{\rm s}=\bigg(q^2{\cal O}(1)+q{\cal O}(\alpha)+{\cal O}(\alpha^2)\bigg)\alpha^{\frac{3}{2}n_{V3}+2n_{V_4}+1} \ , \\
\langle T_{\rm e2}^{ij}\rangle_{\rm us}=\bigg(q^2{\cal O}(1)+q{\cal O}(\alpha)+{\cal O}(\alpha^2)\bigg) \alpha^{\frac{3}{2}n_{V3}+3n_{V_4}+3} \ .
\end{align}
The leading contribution is shown in Fig. 6d.

For the radiative photonic contributions which appear in Fig.\ref{fig:NLObound}, we have following counting rules,
\begin{align}
    \langle T_{\gamma \perp}^{ij}\rangle_{\rm s}=\bigg(q^2{\cal O}(1)+q{\cal O}(\alpha)+{\cal O}(\alpha^2)\bigg)\alpha^{\frac{3}{2}n_{V3}+2n_{V_4}-1} \ , \\
    \langle T_{\gamma \perp}^{ij}\rangle_{\rm us}=\bigg(q^2{\cal O}(1)+q{\cal O}(\alpha^2)+{\cal O}(\alpha^4)\bigg)\alpha^{\frac{3}{2}n_{V3}+3n_{V_4}-2} \ .
\end{align}
Similarly, for the mixed radiative-Coulomb operator, one has
\begin{align}
 \langle T_{\gamma\parallel \perp}^{ij}\rangle_{\rm s}=\bigg(q^2{\cal O}(1)+q{\cal O}(\alpha)+{\cal O}(\alpha^2)\bigg) \alpha^{\frac{3}{2}n_{V3}+2n_{V_4}-\frac{1}{2}} \ , \\
    \langle T_{\gamma\parallel \perp}^{ij}\rangle_{\rm us}=\bigg(q^2{\cal O}(1)+q{\cal O}(\alpha^2)+{\cal O}(\alpha^4)\bigg)\alpha^{\frac{3}{2}n_{V3}+3n_{V_4}-\frac{3}{2}} \ ,
\end{align}
and the same rule applies to $T^{ij}_{\perp p}$ in Eq.~(\ref{eq:mixinter}) as well.

Here we use $T^{ij}_{e0}$ and $T^{ij}_{\gamma \perp}$ as two examples to demonstrate how to derive the above power-counting rule:
\begin{itemize}
    \item The operator insertion itself will contributes to $q^2{\cal O}(1)+q{\cal O}(\alpha)+{\cal O}(\alpha^2)$ in case of $T^{ij}_0$. In case of $T^{ij}_{\gamma \perp}$, one has $q^2{\cal O}(1)+q{\cal O}(\alpha)+{\cal O}(\alpha^2)$ in the soft region while $q^2{\cal O}(1)+q{\cal O}(\alpha^2)+{\cal O}(\alpha^4)$ in the ultra-soft region, which follows from Eq.~(\ref{eq:powerrs}) and Eq.~(\ref{eq:powerrus}).
    \item Each of the electron-photon triple vertices contributes to one power of $\alpha$ from the velocity and half power of $\alpha$ from the interaction, leading to $\alpha^{\frac{3}{2}n_{V_3}}$.
    \item Each of the seagull vertices contributes to one power of $\alpha$ from the interaction, leading to $\alpha^{n_{V_4}}$.
    \item The rest of the diagram, including all the energy-denominators, phase-space measures and momentum integrals has the mass dimension $\lambda^{n_{V_4}}$ for $T^{ij}_{0}$ and $\lambda^{n_{V_4}-1}$ for $T^{ij}_{\rm \gamma \perp}$ with $\lambda=\alpha$ in soft region and $\lambda=\alpha^2$ in ultra-soft region.
\end{itemize}
Combining all the factors leads to the above results.  For non-1PI diagram, it is easy to show that each self-energy-like bubble insertion on external legs will increase at least one factor of $\alpha^2$, depending on the type of insertions.  Finally, diagrams with simultaneous existence of multiple scales will be more suppressed.

Given the above, it is easy to see that in order to obtain order $\alpha$ contribution at $q^2$, one needs the following:
\begin{itemize}
    \item $T_{e2}^{ij}$ with $n_{V_3}=n_{V_4}=0$. This corresponds to Fig.~\ref{fig:fermiontadlamb}. However, we have shown that this diagram is $q$-independent and vanishes in DR.
    \item  $T^{ij}_{\gamma \parallel \perp}$ and $T^{ij}_{\rm \perp p}$ with $n_{V_3}=1,n_{V_4}=0$. They correspond to Fig.~\ref{fig:lambmix} and Fig.~\ref{fig:lambmixinter}, respectively.
    \item  $T^{ij}_{\gamma \perp}$ with $n_{V_3}=2, n_{V_4}=0$. This corresponds to Fig.~\ref{fig:lambradia}.
    \item  $T^{ij}_{\gamma  \perp}$ with $n_{V_4}=1,n_{V_3}=0$. This corresponds to Fig.~\ref{fig:lambtad}.
\end{itemize}
One must notice that for $\langle T^{ij}_{\gamma \parallel \perp}\rangle$ and $\langle T^{ij}_{\perp p}\rangle$, the above power-counting using $|\vec{E}_{\parallel}|_{us}^2 \sim \alpha^9$ leads to ${\cal O}(1)$ at order $q^2$ when $n_{V_3}=1, n_{V_4}=0$ in the ultra-soft region, corresponding to Fig.~\ref{fig:lambmix} and Fig.~\ref{fig:lambmixinter}. However, we will show that in this case $\langle T^{ij}_{\perp p}\rangle=0$, and for $\langle T^{ij}_{\gamma \parallel \perp}\rangle$ one must have $M \ne N$ in Eq.~(\ref{eq:Ecspecial}). Therefore, in this case the actual power-counting should be given by Eq.~(\ref{eq:powercus}), which adds one more $\alpha$, leading to the ${\cal O}(\alpha)$ contribution as well.

In conclusion, one needs to calculate all the photonic contributions in Fig.~\ref{fig:NLObound}. More explicitly, in Fig.~\ref{fig:lambmix} one has the mixed contribution between Coulomb and radiative photon, both emitted from the electron line.  In Fig.~\ref{fig:lambradia} one has a purely radiative contribution, which contributes to order $\alpha$ as well.  Finally, in Fig.~\ref{fig:lambtad}, one has the tadpole contribution. The detailed results are presented in Sec. IV.

\section{Order-$\frac{\alpha}{m_e}$ Matching for $T^{ij}_{\rm NRQED}$}

To calculate the form factor $C$ to the next-to-leading order, we first consider the contribution from the quantum correction to the momentum current, and
match the $T^{ij}_{\rm NRQED}$ to that of QED at order $\alpha$. Since the matching is only sensitive to the UV contribution, it is sufficient to consider a free electron without the background field $V_p$. Furthermore, for our purpose,
we only need to consider spin-independent part.

The matching of the momentum current form factor for the free electron states starts from the full QED result after NR reduction,
\begin{align}\label{eq:QEDform}
    &\langle \vec{p}+\vec{Q}|T^{ij}_{\rm QED}|\vec{p}-\vec{Q}\rangle\nonumber \\
    &=\frac{p^ip^j}{m_e}A(Q^2)+(Q^iQ^j-\delta^{ij}Q^2)\tilde C_{\rm QED}(Q^2)
\end{align}
where $|\vec{p}+\vec{Q}\rangle$ denotes free-electron state with spatial-momentum equals to $\vec{p}+\vec{Q}$. We choose
the Breit frame, and for simplicity define $\vec{Q}\equiv \vec{q}/2$.The EMT form factor $A(Q^2)=1+{\cal O}(\frac{Q^2}{m_e^2})$ receives quantum corrections starting from order $\frac{\alpha Q^2}{m_e^2}$, while $\tilde C_{\rm QED}(Q^2)$ receives corrections at ${\cal O}(\frac{\alpha}{m_e})$.

$T^{ij}_{\rm NRQED}$ may receive corrections beyond the tree-level expression
\begin{align}
    &T^{ij}_{\rm NRQED}=a_{\rm tree}(T^{ij}_{e}+T^{ij}_{\gamma})\nonumber \\
    &+d_0(\partial^{i}\partial^j-\delta^{ij}\partial^2)\Psi^{\dagger}\Psi+{\cal O}(\frac{\alpha}{m_e^2}) \ .
\end{align}
where ${\cal O}(\frac{\alpha}{m_e^2})$ denotes high-dimensional operators such as
\begin{align}
    {\cal O}_{1}^{ij}=\frac{a_1}{m_e^2} \partial^2 T^{ij}_{e0} \ , \\
     {\cal O}_{2}^{ij}=\frac{a_2}{m_e^2} \partial^2(\partial^{i}\partial^j-\delta^{ij}\partial^2)\Psi^{\dagger}\Psi \ ,
\end{align}
and so on, all starting from order $\frac{\alpha}{m_e^2}$. The matching coefficients $a_{\rm tree}$, $d_0$, $a_1$, $a_2$ must be solved in order to reproduce Eq.~(\ref{eq:QEDform}) order by order in $\frac{1}{m_e}$ and $\alpha$:
\begin{align}
    \langle \vec{p}+\vec{Q}|T^{ij}_{\rm QED}|\vec{p}-\vec{Q}\rangle=\langle \vec{p}+\vec{Q}|T^{ij}_{\rm NRQED}|\vec{p}-\vec{Q}\rangle \ .
    \label{eq:matching}
\end{align}
For example, to match to $\frac{Q^2}{m_e^2}$ correction in $A$ one needs $a_1$ and to match to $\frac{Q^2}{m_e^2}$ correction in $\tilde C$ one needs $a_2$. Spin part of the Lagrangian will also be relevant to $a_1$ and $a_2$ as well.

However, as we have already shown in previous section, when the proper power-counting of $p^ip^j \sim {\cal O}(\alpha^2)$ are being taken into account, radiative corrections from $\langle T^{ij}_{e} \rangle_{H}$ will appear only at order $\alpha^3\frac{Q^2}{m_e}$ when averaged in the bound state. The same will apply to $a_1$ and $a_2$. Therefore the only matching constant useful in our calculation is $d_0$, which receives contribution already at order $\alpha$ and is caused purely by $T_{\gamma}^{ij}$. To obtain $d_0$, it is even simpler to work in the frame with $\vec{p}=0$, where the $A$ contribution disappears.

Thus, to compute the matching coefficients using Eq. (  \ref{eq:matching}), we first calculate
\begin{align} \label{eq:formfactor}
    \langle \vec{Q}|T_{\rm tree}^{ij}|-\vec{Q}\rangle=(Q^iQ^j-\delta^{ij}Q^2)\tilde C(Q^2) \ ,
\end{align}
appearing in the right-hand side, and
obtain the explicit formula for $\tilde C(Q^2)$ at order $\frac{\alpha}{m_e}$. We show that it has the same logarithmic divergences when $Q^2\rightarrow 0$ as the full QED, but differs in UV. We perform this calculation in Coulomb gauge with the standard dimensional regularization (DR) with $D=3-2\epsilon$ for UV divergence.

\subsection{Fermionic contributions}

\begin{figure}[h!]
{%
  \includegraphics[height=2cm]{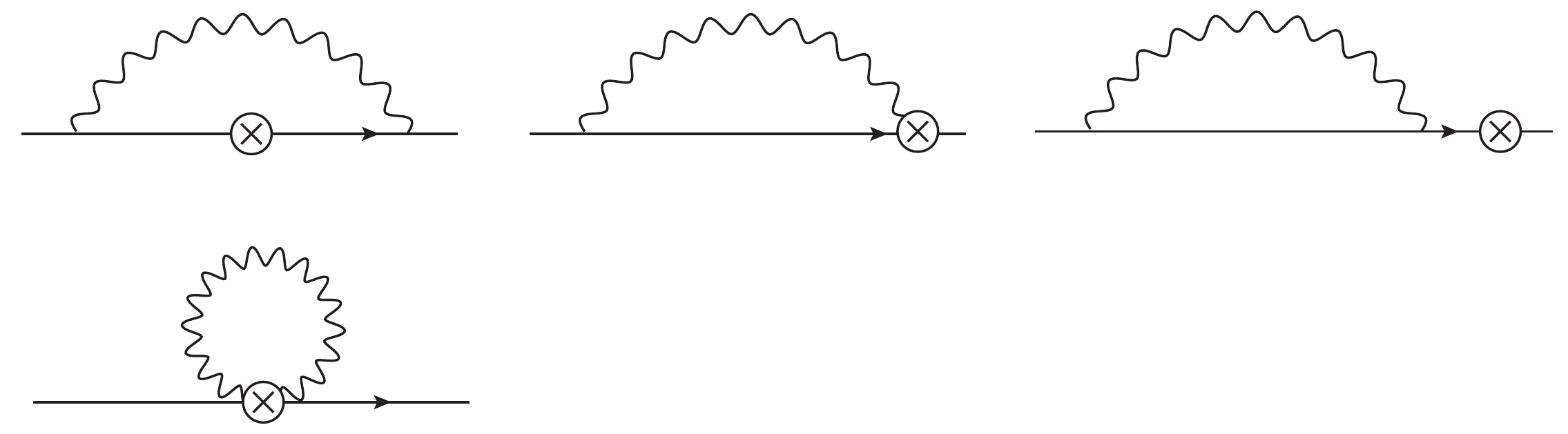}
}
\caption{The contribution of $T^{ij}_{e}$ at one loop. The tadpole diagram, which is absent in the relativistic fermionic theory, is caused by the $\frac{e^2}{m}\Psi^{\dagger}\Psi A^iA^j$ term in $T^{ij}_{e}$. }
\label{fig:singlefermion}
\end{figure}

 The relevant diagrams are shown in Fig~\ref{fig:singlefermion}. To order $\frac{1}{m}$, it is easy to show that in the Coulomb gauge, only the tadpole diagram contributes and reads
\begin{align}\label{eq:fermiontadpole}
    \langle \vec{Q}|T^{ij}_{e}|-\vec{Q}\rangle=\frac{e^2}{2m_e}\mu^{2\epsilon}\int \frac{d^D\vec{k}}{(2\pi)^D|\vec{k}|}\bigg(\delta^{ij}-\frac{k^ik^j}{\vec{k}^2}\bigg) \ .
\end{align}
It is free from IR divergence, $Q$-independent therefore vanishes in DR.  All other diagram are of order at least $\frac{1}{m^2}$ and will not contribute to the matching.

\subsection{Photonic contributions}

\begin{figure}[htb]	
		\begin{subfigure}{.2\textwidth}
		\centering
		\includegraphics[width=\linewidth]{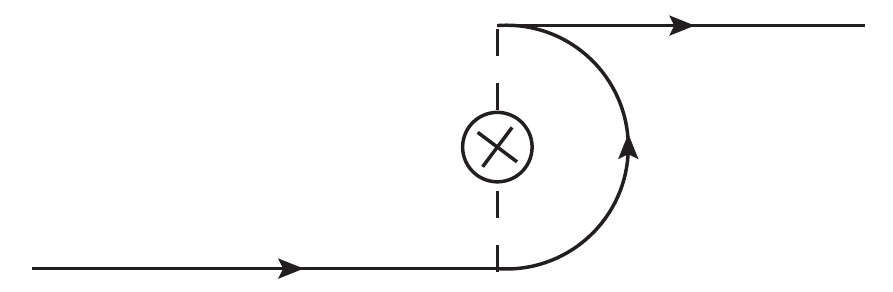}
		\caption{}
		\label{fig:singlec}
	\end{subfigure}
\begin{subfigure}{.29\textwidth}
		\centering
		\includegraphics[width=\linewidth]{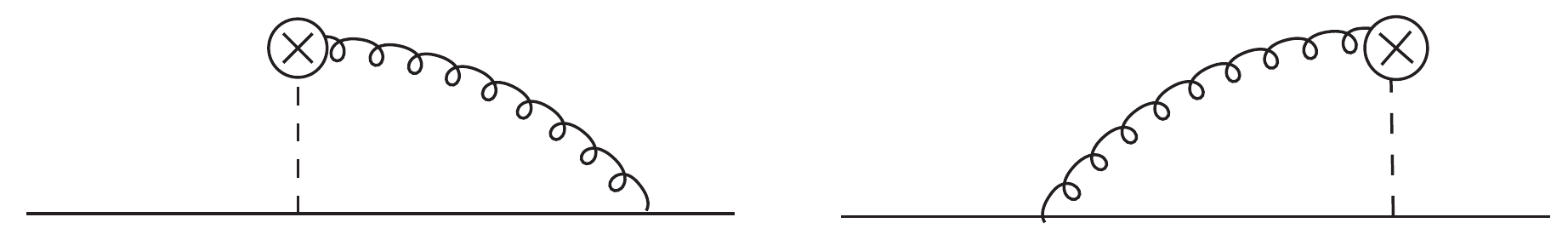}
		\caption{}
		\label{fig:singlemixed}
	\end{subfigure}
 \begin{subfigure}{.2\textwidth}
	\centering
	\includegraphics[width=\linewidth]{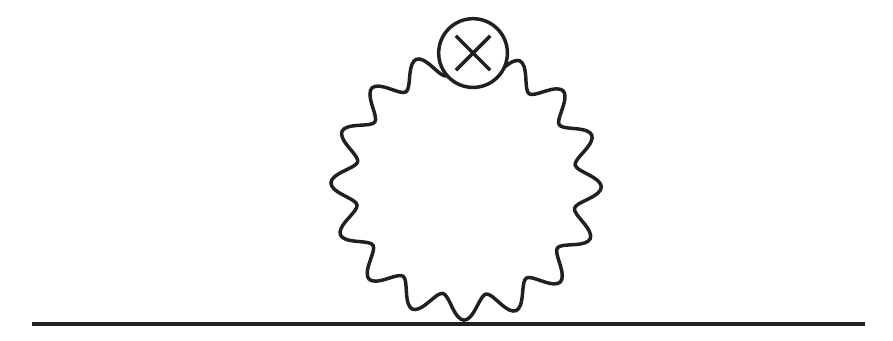}
	\caption{}
	\label{fig:singletad}
\end{subfigure}	
\caption{The contribution of $T_{\gamma}^{ij}$ in non-relativistic reduction. Notice that the tadpole diagram contributes at order $\frac{1}{m_e}$, same as the mixed one.}
		\label{fig:singlenonre}
\end{figure}

One needs to consider the diagrams in Fig.~\ref{fig:singlenonre}. To simplify notation, we write $\langle \vec{Q}|T^{ij}|-\vec{Q}\rangle$ as $\langle T^{ij}\rangle (2\vec{Q})$ and the argument $(2\vec{Q})$ are frequently omitted without causing confusion.

We first consider the pure Coulomb contribution, shown in Fig~\ref{fig:singlec}. This term is transverse by itself and reads
\begin{align}
    \langle T^{ij}_{\gamma \parallel}\rangle_{\rm 9a}(2\vec{Q}) =\frac{\alpha\pi}{8|Q|}(Q^iQ^j-\delta^{ij}Q^2) \ .
\end{align}
The mixed contribution is shown in part b) Fig~\ref{fig:singlemixed}. The two diagrams reads
\begin{align}
    \langle T^{ij}_{\gamma \parallel \perp}\rangle_{\rm 9b}(2\vec{Q})=&e^2\mu^{2\epsilon}\int \frac{d^D\vec{k}}{(2\pi)^D}\frac{(k-Q)^iP^{jk}(\vec{k}+\vec{Q})Q^k+(i\leftrightarrow j)}{m_e|\vec{k}-\vec{Q}|^2|\vec{k}+\vec{Q}|} \nonumber \\
    -&e^2\delta^{ij}\mu^{2\epsilon}\int \frac{d^D\vec{k}}{(2\pi)^D}\frac{(k-Q)^lP^{lk}(\vec{k}+\vec{Q})Q^k}{m_e|\vec{k}-\vec{Q}|^2|\vec{k}+\vec{Q}|} \ ,
\end{align}
where the standard triple vertex $-\frac{ie}{2m_e}(p+p')^j$ and the relation $P^{jk}(\vec{k}+\vec{Q})(Q-k)^k=2P^{jk}(\vec{k}+\vec{Q})Q^k$ have been used.  Due to rotational invariance, one can reduce the above integral to two scalar integrals
\begin{align}\label{eq:7boriginal}
    &\langle T^{ii}_{\gamma \parallel \perp}\rangle_{\rm 9b }(2\vec{Q}) =2e^2(D-2)\mu^{2\epsilon}\int \frac{d^D\vec{k}}{(2\pi)^D}\frac{Q^lP^{lk}(\vec{k}+\vec{Q})Q^k}{m_e|\vec{k}-\vec{Q}|^2|\vec{k}+\vec{Q}|} \ , \\
    &Q^{i}\langle T^{ij}_{\gamma \parallel \perp} \rangle_{\rm 9b}(2\vec{Q}) Q^j=2e^2\mu^{2\epsilon}\int \frac{d^D\vec{k}}{(2\pi)^D}\frac{\vec{k}\cdot \vec{Q}Q^lP^{lk}(\vec{k}+\vec{Q})Q^k}{m_e|\vec{k}-\vec{Q}|^2|\vec{k}+\vec{Q}|} \ .
\end{align}
They are evaluated in Appendix.B. The result in the $\epsilon \rightarrow 0$ limit, reads
\begin{align}\label{eq: 7btranse}
    \langle T^{ii}_{\gamma\parallel\perp}\rangle_{\rm 9b}=\frac{e^2}{3m_e\pi^2 \epsilon}Q^2+\frac{e^2}{3m_e\pi^2}Q^2\bigg(-\ln \frac{Q^2}{\mu^2}+c_1\bigg)  \ ,
\end{align}
and
\begin{align}\label{eq: 7blongi}
    Q^i\langle T^{ij}_{\gamma \parallel \perp}\rangle_{\rm 9b} Q^j=-\frac{e^2}{15m_e\pi^2 \epsilon}Q^4+\frac{e^2}{15m_e\pi^2}Q^4\bigg(\ln \frac{Q^2}{\mu^2}+c_2\bigg)  \ .
\end{align}
with
\begin{align}
    c_1= \gamma_E -5+\ln 4+\ln \pi +2 \psi\left(\frac{5}{2}\right) \ , \\
    c_2=-\gamma_E +3-\ln 4-\ln \pi -2 \psi\left(\frac{7}{2}\right) \ .
\end{align}
Notice that the digamma function is defined as $\psi(z)=\frac{d}{dz}\ln \Gamma(z)$, $\psi(\frac{3}{2})=2-\gamma_E-\ln4$ and one has the recursive relation $\psi(z+1)=\psi(z)+\frac{1}{z}$.
From these, it is clear that the mixed contributions themselves are not transverse.

To obtain a transverse EMT, one must include the contributions from the tadpole diagrams shown in Fig~\ref{fig:singletad} as well. The detail of the calculation is present in Appendix.C, here we only present the result. First, the non-conserved part of Fig.~\ref{fig:singletad} can be calculated as
\begin{align}\label{eq:7nnoncon}
    &Q^i\langle T^{ij}_{\gamma \perp}\rangle_{\rm 9c} Q^j\nonumber \\
    &=e^2Q^4\mu^{2\epsilon}\frac{2^{2 \epsilon-4} (\epsilon-1) \pi ^{\epsilon-\frac{3}{2}} \left(Q^2\right)^{-\epsilon} \Gamma (3-2 \epsilon) \Gamma (\epsilon-1)}{m_e \Gamma \left(\frac{7}{2}-2 \epsilon\right)} \ .
\end{align}
Notice the appearance of $\Gamma(\epsilon-1)$ due to the quadratic divergence. Thus by expanding around  $\epsilon= 0$, one has
\begin{align}\label{eq:7nnonconresult}
    Q^i\langle T^{ij}_{\gamma \perp}\rangle_{\rm 9c} Q^j=\frac{e^2}{15m_e\pi^2\epsilon}Q^4+\frac{e^2}{15m_e\pi^2}Q^4\bigg(-\ln \frac{Q^2}{\mu^2}+c_2'\bigg) \ ,
\end{align}
with
\begin{align}
    c_2'=\gamma_E -3+\ln 4+\ln \pi +2 \psi \left(\frac{7}{2}\right)
\end{align}
Clearly, $c_2=-c_2'$, therefore one has
\begin{align}
    Q^{i}\langle T^{ij}_{\gamma \perp}\rangle_{\rm 9c} Q^j+Q^{i}\langle T^{ij}_{\gamma \perp}\rangle_{\rm 9d} Q^j==0 \ ,
\end{align}
and the $T^{ij}$ is conserved. Similarly, the trace part for Fig.~\ref{fig:singletad} is calculated in Appendix.C as
\begin{align}\label{eq:7dtrace}
     \langle T^{ii}_{\gamma \perp} \rangle_{\rm 9c}(2\vec{Q})=-\frac{Q^2 e^2 c_1'}{18m_e\pi ^2 } \ ,
\end{align}
where
\begin{align}
   c_1'&= 9 \gamma_E -25+3\ln(4)-3 \psi\left(\frac{3}{2}\right)+12 \psi\left(\frac{5}{2}\right)\nonumber \\
   &=1-6\ln 4 \ .
\end{align}
Thus, both of the tadpole and the mixed diagram are required in order to maintain transversity. However, the divergences in the $C(Q)$ can be reads from the $T^{ii}$ for the mixed diagram Fig.\ref{fig:singlemixed} only, while the $T^{ii}$ for the tadpole diagram is logarithm-free.

\subsection{Matching to QED}
We now collect the results and match to full QED.  By combining Eq.~(\ref{eq: 7btranse}) and Eq.~(\ref{eq:7dtrace}), the full contribution of $T^{ij}_{\rm tree}$ reads
\begin{align}
    \langle T^{ij}_{\rm tree}\rangle (2\vec{Q})=(Q^iQ^j-\delta^{ij}Q^2)\tilde C(Q^2) \ ,
\end{align}
where
\begin{align}
    \tilde C(Q^2)=\frac{\alpha\pi}{8|Q|}+\frac{e^2}{6m_e\pi^2}\bigg(-\frac{1}{\epsilon}+\ln \frac{Q^2}{\mu^2}+\gamma_E-\ln\pi-\frac{7}{6}\bigg)  \ .
\end{align}
In comparison,  the small-$Q$ asymptotics of $C$-form factor for relativistic electron in the full QED can be obtained from literature~\cite{Berends:1975ah,Milton:1977je} as
\begin{align}\label{eq:CQED}
    \tilde C_{\rm QED}(Q^2)=\frac{\alpha\pi}{8|Q|}+\frac{e^2}{6m_e\pi^2 }\ln \frac{4Q^2}{m_e^2}-\frac{11e^2}{72m_e\pi^2 }
\end{align}
It has the same IR structure as the NRQED, but differs in UV. The required formulas are collected in Appendix.G.

It is clear now that in order to match to the full QED, one simply needs to add to the tree-level
momentum current of the NRQED the following local counter term
\begin{align}
    T^{ij}_{\rm NRQED}= T^{ij}_{\rm tree}+d_0(\partial^i\partial^j-\delta^{ij}\partial^2)\Psi^{\dagger}\Psi \ ,
\end{align}
where $d_0$ contain only logarithms in $\mu$ and $m_e$,
\begin{align}
    d_0=-\frac{\alpha}{6\pi m_e}\bigg(\frac{1}{\epsilon}+\ln \frac{4\mu^2}{m_e^2}+\ln \pi -\gamma_E+\frac{1}{4}\bigg) \ .
\end{align}
This concludes our construction of the momentum current $T^{ij}$ in NRQED.

\section{${\cal O}(\alpha)$ Radiative Corrections to $\tau_H$}

After obtaining the matching coefficient $d_0$, in this section we calculate the $\langle 0|T^{ij}_{\rm NRQED} |0\rangle_H$ in the bound-state and obtain the final result for $\tau_H$.  More explicitly, the calculation proceeds as follows.

We will present the result for the mixed diagrams Fig.~\ref{fig:lambmix} and Fig.~\ref{fig:lambmixinter} and then for the radiative diagram Fig.~\ref{fig:lambradia}. We show explicitly that the non-conserved part $Q^{i}T^{ij}Q^j$ cancel with the tadpole diagram Fig.~\ref{fig:lambtad} that is essentially independent of the bound state and remains the same as the free NRQED calculation in previous section. The power-counting rules in the ultra-soft region are used to decouple the matrix elements and the momentum integrals, therefore in principle our calculation is only valid in this region as well. However, it is not hard to show that in the the mixed diagram, the only diagram which diverges in UV, our formulas hold in the soft region as well. Therefore, the result has the same UV structure as the free NRQED and can be matched to the full QED using the same matching coefficient $d_0$ obtained in the previous section, which leads to our final result Eq.~(\ref{eq:finalre}). Its numerical value will also be provided.

For notational simplicity, the momentum transfer will be $\vec{q}=2\vec{Q}$ and we will use the following notation frequently
\begin{align}
    \langle T^{ij}\rangle (2\vec{Q})\equiv \int d^D \vec{x}e^{-i2\vec{Q}\cdot \vec{x}}\langle 0|T^{ij}(\vec{x})|0\rangle \ .
\end{align}
Without causing confusion, the argument $(2\vec{Q})$ will be omitted. Since we are only interested in the small-$Q$ behavior of the form factor, expansion to quadratic order in $Q$ will always be understood.

\subsection{The mixed diagrams Fig.~\ref{fig:lambmix} and Fig.~\ref{fig:lambmixinter}}
We first consider the mixed diagrams in Fig.~\ref{fig:lambmix} and Fig.~\ref{fig:lambmixinter}. We show that the interference diagram Fig.~\ref{fig:lambmixinter} vanishes. Indeed, using the Feynman rule, one has
\begin{align}
    &\langle T^{ij}_{\perp p}\rangle_{\rm 7b} =-e^2\mu^{2\epsilon}\int\frac{d^D\vec{k}}{(2\pi)^D}\frac{(k-2Q)^i}{|\vec{k}-2\vec{Q}|^2}\frac{P^{jl}(\vec{k}) v^l_{00}(\vec{k})}{|\vec{k}|}\nonumber \\
    &+(i\leftrightarrow j)+\delta^{ij}e^2\mu^{2\epsilon}\int\frac{d^D\vec{k}}{(2\pi)^D}\frac{Q^l}{|\vec{k}-2\vec{Q}|^2}\frac{P^{ll'}(\vec{k}) v^{l'}_{00}(\vec{k})}{|\vec{k}|}\ ,
\end{align}
which vanishes due to the fact that
\begin{align}
   \vec{v}_{00}(\vec{k})= \int d^D\vec{x} \psi^{\dagger}_{0}(\vec{x})\frac{-i\nabla }{m_e}\psi_{0}(\vec{x})e^{-i\vec{k}\cdot \vec{x}}\propto \vec{k}
\end{align}
which contracts to zero with $P^{ij}(\vec{k})$.

Therefore, it remains to calculate Fig.~\ref{fig:lambmix}.  We start with $T^{ii}$. Notice that for $k={\cal O}(\alpha^2 m_e)$, one has the standard dipole-expansion~\cite{Weinberg:1995mt,Pineda:1997ie} of the matrix elements defined in Eq.~(\ref{eq:matrix})
\begin{align}
   v^{i}(\vec{k})= v^i_{MN} +{\cal O}(\alpha)\ , \\
    \rho_{MN}(\vec{k}) =\delta_{MN}-i\vec{k}\cdot \vec{x}_{MN} +{\cal O}(\alpha) \ .
\end{align}
where $\vec{v}=-\frac{i\vec{\nabla}}{m_e}$. Using these one has for the trace part
\begin{align}\label{eq:5atrace}
   &\langle T^{ii}_{\gamma \parallel \perp}\rangle_{\rm 7a} = -2(D-2)\mu^{2\epsilon}\frac{e^2}{D}\sum_{M}i(x^i_{0M}v^i_{M0}-v^i_{0M}x^i_{M0})\nonumber \\
   &\times \int \frac{d^Dk}{(2\pi)^D}\frac{Q^lP^{lk}(\vec{k})Q^k}{|\vec{k}-2\vec{Q}|^2\bigg(|\vec{k}|+E_M-E_0\bigg)}  \ .
\end{align}
Clearly, the divergent part is independent of the bound-state thanks to canonical commutation relation $[x,p]=i$ in any dimension
\begin{align}
-\frac{1}{D}\sum_{M}i(x^i_{0M}v^i_{M0}-v^i_{0M}x^i_{M0})\equiv \frac{1}{m_e}\ .
\end{align}
In particular, in the soft region where $ |\vec{k}|={\cal O}(\alpha m_e)$, the formula above is also valid at small $Q$, after neglecting the binding energies in the denominator. The integral is calculated in Appendix.D. The result reads
\begin{widetext}
\begin{align}\label{eq:5aresult1}
  \langle T^{ii}_{\gamma \parallel \perp}\rangle_{\rm 7a} (2\vec{Q})= \frac{e^2}{\pi^2D}Q^2\sum_{M}i(\vec{v}_{0M}\cdot \vec{x}_{M0}-\vec{x}_{0M}\cdot \vec{v}_{M0})\bigg(\frac{1}{3\epsilon}-\frac{1}{3}\ln \frac{(E_M-E_0)^2}{\mu^2}+\frac{-3\gamma_E+3\ln \pi -1}{9} \bigg)
\end{align}
At this step, it is helpful to perform certain simplification of the matrix elements. Notice the following equalities in $D$ dimensions
\begin{align}
    \langle N|[H,\vec{x}]|M\rangle=-i\langle N|\vec{v}|M\rangle=(E_N-E_M)\langle N|\vec{x}|M \rangle \ ,
\end{align}
thus for any $M \ne N$ \ ,
\begin{align}
   - i (\vec{x}_{NM}\cdot \vec{v}_{MN}-\vec{v}_{NM}\cdot\vec{x}_{MN})=2\frac{\vec{v}_{NM}\cdot \vec{v}_{MN}}{E_M-E_N} \ .
\end{align}
Therefore one has
\begin{align}
      \langle T^{ii}_{\gamma \parallel\perp}\rangle_{\rm 7a} (2\vec{Q})= \frac{e^2}{\pi^2}Q^2\sum_{M}\frac{2\vec{v}_{M0}\cdot \vec{v}_{0M}}{D(E_M-E_0)} \bigg(\frac{1}{3\epsilon}-\frac{1}{3}\ln \frac{(E_M-E_0)^2}{\mu^2}+\frac{-3\gamma_E+3\ln \pi -1}{9} \bigg)  \ .
\end{align}
in which the matrix elements $\vec{x}_{MN}$ are eliminated.

Similarly, the non-conserved part $Q^i\langle T^{ij}_{\gamma \parallel \perp}\rangle_{\rm 7a} Q^j$  can be calculated as
\begin{align}\label{eq:5anoncon}
    Q^i \langle T^{ij}_{\gamma \parallel \perp} \rangle_{\rm 7a} Q^j= \sum_{M}\frac{2\vec{v}_{M0}\cdot \vec{v}_{0M}}{D(E_M-E_0)}2e^2\mu^{2\epsilon}\int \frac{d^D\vec{k}}{(2\pi)^D}\frac{(\vec{k}\cdot \vec{Q}-\vec{Q}^2)(Q^2\vec{k}^2-(\vec{k}\cdot \vec{Q})^2)}{|\vec{k}-2\vec{Q}|^2|\vec{k}|^2(|\vec{k}|+E_M-E_0)}  \ .
\end{align}
By parameterizing as usual, one obtains
\begin{align}\label{eq:5anonconresult}
    Q^i \langle T^{ij}_{\gamma \parallel \perp} \rangle_{\rm 7a} Q^j= \frac{Q^4e^2}{\pi^2}\sum_{M}\frac{2\vec{v}_{M0}\cdot \vec{v}_{0M}}{D(E_M-E_0)} \bigg(-\frac{1}{15\epsilon}+\frac{1}{15}\ln \frac{(E_M-E_0)^2}{\mu^2}+\frac{-15\ln \pi+15\gamma_E-1}{225}\bigg) \ .
\end{align}
The detail of the calculation is presented in Appendix.D. It contains the same UV divergence as the mixed diagram for the single electron, but the IR divergence $\ln Q^2$ is regulated by the binding energy differences. Therefore, to cancel the $\ln Q^2$ in the tadpole diagram, there must be contributions of the form $\ln \frac{Q^2}{(E_M-E_0)^2}$. As we will see, the radiative diagram Fig.~\ref{fig:lambradia} will exactly produce this missing piece.

\subsection{Purely radiative contribution Fig.~\ref{fig:lambradia}}
The last but the most complicate diagram that remains to be calculated is the pure radiative contribution in Fig.~\ref{fig:lambradia}. We start with the $\langle T^{ii}_{\gamma \perp}\rangle_{\rm 7c}$. Using standard Feynman rule, this can be written as
\begin{align}\label{eq:radiab}
    \langle T^{ii}_{\gamma \perp}\rangle_{\rm 7c}=\sum_{M}\vec{v}_{0M}\cdot \vec{v}_{M0}\frac{2ie^2}{D}\mu^{2\epsilon}\int \frac{dk^0d^{D}\vec{k}}{(2\pi)^{D+1}}\frac{A(k,Q,D){\rm tr}(P(k-Q)P(k+Q))+2(4-D)Q^TP(k-Q)P(k+Q)Q}{\bigg(k_0^2-(\vec{k}-\vec{Q})^2+i0\bigg)\bigg(k_0^2-(\vec{k}+\vec{Q})^2+i0\bigg)\bigg(E_0-k^0-E_M+i0\bigg)} \ ,
\end{align}
where $A(k,Q,D)$ is defined in Eq.~(\ref{eq:AkQD}). Notice the similarity of the integrand to the tadpole contribution in Eq.~(\ref{eq:5d}). Since the calculation is rather tedious, we present all the details in Appendix.E. The result reads
\begin{align}\label{eq:5cresult}
     \langle T^{ii}_{\gamma \perp}\rangle_{\rm 7c} (2\vec{Q})=-Q^2\frac{2e^2\ln4}{3\pi^2}\sum_{M}\frac{\vec{v}_{0M}\cdot \vec{v}_{M0}}{D(E_M-E_0)} \ .
\end{align}
We then move to the non-conserved part $Q^i\langle T^{ij}_{\gamma \perp}\rangle_{\rm 7c} Q^j$. This is the most involved part of the calculation and is presented in Appendix.E. The result reads
\begin{align}\label{eq:5cnonresult}
    Q^i \langle T^{ij}_{\gamma \perp}\rangle_{\rm 7c} Q^j= \frac{Q^4e^2}{\pi^2}\sum_{M}\frac{2\vec{v}_{0M}\cdot \vec{v}_{M0}}{D(E_M-E_0)}\bigg(\frac{1}{15}\ln \frac{Q^2}{(E_M-E_0)^2}+\frac{15\ln 4-46}{225}\bigg) \ .
\end{align}
As claimed in the previous subsection, it contains the missing  $\ln \frac{Q^2}{(E_0-E_M)^2}$ with the correct coefficient.

\subsection{Checking conservation.  }
After finishing the difficult part of the calculation, here we check the conservation of $T^{ij}$. Combining Eq.~(\ref{eq:5anonconresult}) and Eq.~(\ref{eq:5cnonresult}), the $\ln (E_M-E_0)^2$ cancels, left with
\begin{align}\label{eq:Tlongira}
    Q^i\left(\langle T^{ij}_{\gamma \parallel \perp}\rangle_{\rm 7a} +\langle T^{ij}_{\gamma \perp}\rangle_{\rm 7c} \right)Q^j=\frac{Q^4e^2}{\pi^2}\sum_{M}\frac{2\vec{v}_{M0}\cdot \vec{v}_{0M}}{D(E_M-E_0)} \bigg(-\frac{1}{15\epsilon}+\frac{1}{15}\ln \frac{Q^2}{\mu^2}+\frac{15\ln4-15\ln \pi+15\gamma_E-47}{225}\bigg)  \ ,
\end{align}
which can be simplified after using the sum-rule Eq.~(\ref{eq:sumrule}) as
\begin{align}\label{eq:Tlongirare}
     Q^i\left(\langle T^{ij}_{\gamma \parallel \perp}\rangle_{\rm 7a} +\langle T^{ij}_{\gamma \perp}\rangle_{\rm 7c}\right)Q^j=\frac{Q^4e^2}{m_e\pi^2} \bigg(-\frac{1}{15\epsilon}+\frac{1}{15}\ln \frac{Q^2}{\mu^2}+\frac{15\ln4-15\ln \pi+15\gamma_E-47}{225}\bigg)  \ .
\end{align}
Here we show that it cancels with the tadpole contribution in Fig.~\ref{fig:lambtad}. Indeed, since the tadpole contribution is essentially independent of the bound-state, one has the same expression at small $Q$ as Eq.~(\ref{eq:7nnonconresult})
\begin{align}
    Q^i\langle T^{ij}_{\gamma \perp}\rangle_{\rm 7d} Q^j=\frac{e^2}{15m_e\pi^2\epsilon}Q^4+\frac{e^2}{15m_e\pi^2}Q^4\bigg(-\ln \frac{Q^2}{\mu^2}+c_2'\bigg) \ ,
\end{align}
where
\begin{align}
    c_2'=\gamma_E -3+\ln 4+\ln \pi +2 \psi \left(\frac{7}{2}\right)\equiv \frac{47}{15}-\gamma_E +\ln \pi- \ln 4 \ .
\end{align}
To obtain this we used again the well-known relation for digamma function
\begin{align}
    \psi(n+\frac{1}{2})=-\gamma_E-\ln 4+\sum_{k=1}^n \frac{2}{2k-1} \ .
\end{align}
Therefore, for the bound state we have shown that the momentum current at order $\frac{\alpha}{m}$ is purely transverse
\begin{align}
    Q^i\left(\langle T^{ij}_{\gamma \parallel \perp}\rangle_{\rm 7a}+\langle T^{ij}_{\gamma \perp}\rangle_{\rm 7c+7d}\right)Q^j\equiv 0 \ .
\end{align}
This is the most crucial consistency check of the whole calculation.
\subsection{The total result}
After showing the conservation of $T^{ij}$, we collect all the pieces of $T^{ii}$ and obtain the final result.
First, for the mixed diagram Fig~\ref{fig:lambmix}, one has
\begin{align}
  \langle T^{ii}_{\gamma \parallel\perp}\rangle_{\rm 7a}(2\vec{Q})= \frac{e^2}{\pi^2}Q^2\sum_{M}\frac{2\vec{v}_{M0}\cdot \vec{v}_{0M}}{D(E_M-E_0)} \bigg(\frac{1}{3\epsilon}-\frac{1}{3}\ln \frac{(E_M-E_0)^2}{\mu^2}+\frac{-3\gamma_E+3\ln \pi -1}{9} \bigg)  \ .
\end{align}
For the radiative part Fig.~\ref{fig:lambradia}, one has
\begin{align}
    \langle T^{ii}_{\gamma \perp}\rangle_{\rm 7c} (2\vec{Q})=-\frac{e^2}{\pi^2}Q^2\sum_{M}\frac{2\vec{v}_{0M}\cdot \vec{v}_{M0}}{D(E_M-E_0)}\frac{\ln4}{3}  \ .
\end{align}
For the tadpole part Fig.~\ref{fig:lambtad} , one has the same small-$Q$ result as in Eq.~(\ref{eq:7dtrace})
\begin{align}
   \langle T^{ii}_{\gamma \perp}\rangle_{\rm 7d} (2\vec{Q})=-\frac{e^2}{\pi^2}Q^2\sum_{M}\frac{2\vec{v}_{0M}\cdot \vec{v}_{M0}}{D(E_M-E_0)}\frac{1-6\ln 4}{18} \ .
\end{align}
Therefore, combining all them, one has
\begin{align}
    \langle T_{\rm tree}^{ii}\rangle (2\vec{Q})=\frac{e^2}{\pi^2}Q^2\sum_{M}\frac{2\vec{v}_{M0}\cdot \vec{v}_{0M}}{D(E_M-E_0)} \bigg(\frac{1}{3\epsilon}-\frac{1}{3}\ln \frac{(E_M-E_0)^2}{\mu^2}+\frac{-\gamma_E+\ln \pi }{3}-\frac{1}{6} \bigg) \ ,
\end{align}
which leads to
\begin{align}
    \langle T^{ij}_{\rm tree}\rangle (2\vec{Q})=(Q^iQ^j-\delta^{ij}Q^2)\tilde C_{\rm s}(Q^2) \ ,
\end{align}
where
\begin{align}
    \tilde C_{\rm s}(Q^2=0)=\frac{e^2}{6\pi^2}\sum_{M}\frac{2\vec{v}_{M0}\cdot \vec{v}_{0M}}{D(E_M-E_0)} \bigg(-\frac{1}{\epsilon}+\ln \frac{(E_M-E_0)^2}{\mu^2}+\gamma_E-\ln \pi-\frac{1}{2} \bigg) \ .
\end{align}
To match it to QED, one only needs to add to the above result $-4d_0$
\begin{align}
    \tilde C(Q^2=0)=\tilde C_{\rm s}(Q^2=0)-4d_0=\frac{e^2}{6\pi^2}\sum_{M}\frac{2\vec{v}_{M0}\cdot \vec{v}_{0M}}{D(E_M-E_0)} \bigg(\ln \frac{4(E_M-E_0)^2}{m_e^2}-\frac{1}{4} \bigg)  \ .
\end{align}
Since our $\vec{Q}$ is twice of the momentum transfer $\vec{q}=2\vec{Q}$, one finally has
\begin{align}
    \langle T^{ij}\rangle_H (\vec{q})=(q^iq^j-\delta^{ij}q^2)\frac{C_H(q)}{m_e}\ ,
\end{align}
with
\begin{align}\label{eq:finalre}
\tau_H=\frac{C_H(0)}{m_e}=\frac{1}{4m_e}+\frac{\alpha}{6\pi}\sum_{M}\frac{2\vec{v}_{M0}\cdot \vec{v}_{0M}}{D(E_M-E_0)} \bigg(\ln \frac{4(E_M-E_0)^2}{m_e^2}-\frac{1}{4} \bigg)  \ .
\end{align}
This is the major result of the paper.  Notice that the leading order result has been added.
\end{widetext}
To estimate how large the order $\alpha$ contribution is, one needs to calculate the sum over $M$. If $E_M-E_0$ is in the numerator, this is called the Bethe logarithm and receives large contribution from the continuum spectrum. In our case, we expect the continuum spectrum is also important. In fact, after re-scaling, the contribution can be written as
\begin{align}
    \tau_H=\frac{1}{4m_e}+\frac{\alpha}{6\pi m_e} \bigg(\ln \alpha^4+\tau_{\rm d}+\tau_{\rm c}-\frac{1}{4}\bigg) \ ,
\end{align}
where $\tau_d=-0.264$ and $\tau_c=0.458$ are contributions from the discrete and continuum spectrum, which are defined and evaluated in Appendix F. Put in numbers, one has
\begin{align}\label{eq:resultnu}
    \frac{\tau_H}{\tau_0}-1=\frac{4\alpha}{3\pi } \bigg(\ln \alpha^2-0.028\bigg)=-3.07\times 10^{-2} \ .
\end{align}
Although opposite in sign, the order $\alpha$ contribution is two orders of magnitude smaller comparing to the leading order contribution.

\section{Comment and Conclusion}
Before ending the paper, here we briefly comment on the sign of $\tau_H$. One first notice that in the result Eq.~(\ref{eq:resultnu}), $\ln \alpha^2$ dominate over the constant $-0.056$, therefore the sign at order $\alpha$ is mainly due to the logarithms, which already appears at the level of single electron. In fact, from the calculation we have learned that only the mixed diagram contributes to this logarithm, while the purely-radiative and tadpole diagrams contribute only to the constant.

Besides our calculation in NRQED with dimensional regularization, one can also perform the calculation directly from the dressed Dirac theory in a way similar to Ref.~\cite{Weinberg:1995mt}. In order to obtain the correct expansion in $\alpha$, one should separate the high-energy and low energy contributions into two parts
\begin{align}
    \frac{1}{k^2+i0}\rightarrow \frac{1}{k^2-\mu^2+i0}+\bigg(\frac{1}{k^2+i0}-\frac{1}{k^2-\mu^2+i0}\bigg) \ ,
\end{align}
with the fictitious photon mass $\mu$ satisfying $\alpha^2m_e \ll \mu \ll \alpha m_e$. In the first term, the photon mass will guarantee that the ultra-soft region is non-essential, and the calculation can be performed by completely neglecting the bound-state structure for a single relativistic electron. The second term can be calculated using non-relativistic approximations for the electron as usual, with $\mu$ playing the role of the UV regulator. In fact, one may think that the first term just defines the ``matching constant'' $d_0$ in this scheme. The trouble with the photon mass regulator is that the EMT is not guaranteed to be conserved for finite $\mu$, and the power-divergences in $\mu$ requires additional attention. On contrary, the EMT in dimensional regularization is automatically conserved, and the power-divergence disappears in DR as well.

In conclusion, we have constructed the momentum current density of NRQED up to order $\frac{1}{m_e}$, from which the ${\cal O}(\alpha)$ tensor monopole moment $\tau_H$ for the ground state of hydrogen atom is calculated. Although suffering from IR divergence for a single free electron, $\tau_H$ is finite and remains positive after including the ${\cal O}(\alpha)$ correction. The IR logarithm in NRQED is naturally regulated by the binding energy differences, and the fictitious UV divergence of NRQED ``matches'' precisely with the IR divergence of the relativistic theory, guarantee the ultimate consistency of our calculation. The final
result is similar in expression to the famous Lamb shift of the energy levels.

{\it Acknowledgment.}---
This research is supported by the U.S. Department of Energy, Office of Science, Office of Nuclear Physics, under contract number DE-SC0020682, and by the Priority Research Area SciMat under the program Excellence Initiative - Research University at the Jagiellonian University in Krak\'{o}w.

\appendix
\section{Conservation of $T^{ij}_{\rm tree}$ }
In this appendix we show that $T^{ij}_{\rm tree}$ is conserved. Indeed, using the equation of motion and commutators of $D^i,D^j$
\begin{align}
    (iD^0+\frac{D^iD^i}{2m_e})\Psi=0 \ , \\
    [D^iD^i,D^j]=-2ieF^{ij}D^i-ie\partial^iF^{ij} \ ,
\end{align}
it is not hard to show that
\begin{align}\label{eq:fermionT}
    \partial^i T^{ij}_{e}=&\frac{i}{2}\partial_0(\Psi^{\dagger}D^j\Psi-(D^j\Psi)^{\dagger}\Psi)\nonumber \\ &-F^{j0}J_0-F^{ji}J_i+e\partial^jV_p\Psi^{\dagger}\Psi \ ,
\end{align}
where the last term is due to the the static-potential $V_p$ in $D^0=\partial_0-ieA^0-ieV_p$, and with the electric current reads
\begin{align}
    &J^0=-e\Psi^{\dagger}\Psi \ , \\
    &J^i=-\frac{ie}{2m_e}\bigg(\Psi^{\dagger}D^i\Psi-(D^i\Psi)^{\dagger}\Psi\bigg) \ .
\end{align}
Using the equation of motion for the electric-magnetic field and the Bianchi identity, it is easy to show that
\begin{align}
\partial_iT_{\gamma}^{ij} =-\partial_0 T_{\gamma}^{0j}-F^{j0}J_0-F^{ji}J_i \ ,
\end{align}
 Therefore, up to time derivatives, one has
\begin{align}
    \partial^i \bigg(T^{ij}_{e}+T^{ij}_{\gamma}+T_{\parallel p}^{ij}\bigg)=ie\int \frac{d^3\vec{p}}{(2\pi)^3}p^jV_e(\vec{p}) \ .
\end{align}
For the spherical symmetric ground state $V_e(\vec{p})=V_e(|p|)$, the above normally integrate to zero, therefore implies the current conservation. Similarly, using the transversal condition and the fact that $V_p$ is time independent, one has
\begin{align}
    \partial^i T^{ij}_{\perp p}=\partial^0\bigg(\partial^iV_pF^{ij}+\nabla^2 V_p A^j\bigg) \ ,
\end{align}
which vanishes in energy eigenstates. We will show in Sec.IV that this term vanishes identically for the ground state. For higher exited states this term should be included.

\section{Calculation of Eq.~(\ref{eq:7boriginal})}
\begin{widetext}
Using the standard Feynman-type parametrization one has
\begin{align}
    &\langle T^{ii}_{\gamma \parallel \perp}\rangle_{\rm 9b}=2(D-2)\frac{e^2}{m_eI_1}Q^2(1-\frac{1}{D})\frac{D}{2}\bigg(\frac{1}{4\pi}\bigg)^{\frac{D}{2}}\mu^{2\epsilon}\int_{0}^1 dxx^{\frac{1}{2}}\int_{0}^{\infty} d\rho \rho^{-1+\epsilon}\exp \bigg[-4Q^2\rho x(1-x)\bigg] \ ,
    \end{align}
and
\begin{align}
    &Q^i\langle T^{ij}_{\gamma \parallel \perp}\rangle_{\rm 9b} Q^j=-2\frac{e^2}{m_eI_1}Q^4(1-\frac{1}{D})\frac{D}{2}\bigg(\frac{1}{4\pi}\bigg)^{\frac{D}{2}}\mu^{2\epsilon}\int_{0}^1 dxx^{\frac{1}{2}}(2x-1)\int_{0}^{\infty} d\rho \rho^{-1+\epsilon}\exp \bigg[-4Q^2\rho x(1-x)\bigg] \ ,
    \end{align}
    with $I_1=2\int_{0}^{\infty} dx x^2e^{-x^2}=\frac{\sqrt{\pi }}{2}$. Clearly, there is only logarithmic UV divergence, but not any power-divergence. Performing the integrals and then take the $\epsilon \rightarrow 0$ limit, one reproduces Eq.~(\ref{eq: 7btranse}) and Eq.~(\ref{eq: 7blongi}).

 \section{Calculation of Fig.~\ref{fig:singletad}}
In this appendix we calculate the tadpole diagram shown in Fig.~\ref{fig:singletad}. For this term one needs the electric and magnetic parts of $T^{ij}_{\rm \perp}$ defined in Eq.~(\ref{eq:TE}) and Eq.~(\ref{eq:TB}). One also needs the relations for the projections
\begin{align}
    P^{ij}(\vec{k})P^{ji}(\vec{k}+2\vec{Q})=D-1-4\frac{k^2Q^2-(\vec{k}\cdot \vec{Q})^2}{|\vec{k}|^2|\vec{k}+\vec{2Q}|^2} \ , \\
    Q^TP(\vec{k})P(\vec{k}+2\vec{Q})Q=\frac{k^2Q^2-(\vec{k}\cdot \vec{Q})^2}{|\vec{k}|^2|\vec{k}+\vec{2Q}|^2}(Q^2+2\vec{k}\cdot \vec{Q}) \ ,
\end{align}
which express these scalar functions in terms of projection operator along $\vec{Q}$. We first calculate the non-conserved part, by using the standard Feynman rules one has

 \begin{align}
    Q^i\langle T^{ij}_{\gamma E}\rangle_{\rm 9c} Q^j= \frac{ie^2}{2m_e}\mu^{2\epsilon}\int \frac{d^{D+1}k}{(2\pi)^{D+1}}\frac{k_0^2\bigg[Q^2{\rm tr}P(k-Q)P(k+Q)-2Q^TP(k-Q)P(k+Q)Q\bigg]}{\bigg(k_0^2-(\vec{k}-\vec{Q})^2+i0\bigg)\bigg(k_0^2-(\vec{k}+\vec{Q})^2+i0\bigg)}\ ,
\end{align}
and
\begin{align}
    Q^i\langle T^{ij}_{\gamma B}\rangle_{\rm 9c} Q^j=\frac{ie^2}{2m_e}\mu^{2\epsilon}\int \frac{d^{D+1}k}{(2\pi)^{D+1}}\frac{{\rm tr}P(k-Q)P(k+Q)A(k,Q)+Q^TP(k-Q)P(k+Q)QB(k,Q)}{\bigg(k_0^2-(\vec{k}-\vec{Q})^2+i0\bigg)\bigg(k_0^2-(\vec{k}+\vec{Q})^2+i0\bigg)} \ ,
\end{align}
where $d^{D+1}k\equiv dk^0d^D\vec{k}$, and
\begin{align} \label{eq:ABdef}
    A(k,Q)=\vec{Q}\cdot(\vec{k}-\vec{Q})\vec{Q}\cdot(\vec{k}+\vec{Q})-\frac{Q^2}{2}(\vec{k}-\vec{Q})\cdot(\vec{k}+\vec{Q}) \ , \\
    B(k,Q)=(\vec{k}-\vec{Q})\cdot(\vec{k}+\vec{Q})-4\vec{Q}\cdot(\vec{k}-\vec{Q})-2Q^2 \ .
\end{align}
By combining them, one obtains for $D=3-2\epsilon$
\begin{align}
    Q^i\langle T^{ij}_{\gamma \perp}\rangle_{\rm 9c} Q^j=&\frac{e^2}{2m_e}Q^2\mu^{2\epsilon}\int \frac{d^D\vec{k}}{(2\pi)^D}\frac{1-\epsilon}{|\vec{k}|}-\frac{e^2}{2m_e}\mu^{2\epsilon}\int \frac{d^D\vec{k}}{(2\pi)^D}\frac{\vec{k}^2\vec{Q}^2-(\vec{k}\cdot \vec{Q})^2}{|\vec k|^2|\vec{k}+2\vec{Q}|^3}(\vec{k}^2+2\vec{k}\cdot \vec{Q}+2Q^2) \nonumber \\
    &+\frac{2ie^2}{m_e}\mu^{2\epsilon}\int \frac{d^{D+1}k}{(2\pi)^{D+1}}\frac{(1-\epsilon)(\vec{k}\cdot \vec{Q})^2}{\bigg(k_0^2-(\vec{k}-\vec{Q})^2+i0\bigg)\bigg(k_0^2-(\vec{k}+\vec{Q})^2+i0\bigg)} \nonumber \\
    &-\frac{2ie^2}{m_e}\mu^{2\epsilon}\int \frac{d^{D+1}k}{(2\pi)^{D+1}}\frac{\vec{Q}^2+\vec{k}\cdot \vec{Q}}{\bigg(k_0^2-\vec{k}^2+i0\bigg)\bigg(k_0^2-(\vec{k}+2\vec{Q})^2+i0\bigg)}\frac{\vec{k}^2\vec{Q}^2-(\vec{k}\cdot \vec{Q})^2}{|\vec{k}|^2} \ .
\end{align}
Parameterizing these integrals as usual, one has
\begin{align}
    Q^i\langle T^{ij}_{\gamma \perp}\rangle_{\rm 9c} Q^j=\frac{e^2}{m_e\sqrt{\pi}}(4\pi)^{-\frac{D}{2}}\mu^{2\epsilon}\int_{0}^1 dx\int_{0}^{\infty}d\rho \rho^{-\frac{D}{2}+\frac{1}{2}}e^{-4Q^2\rho x(1-x)}\bigg(Q^2\rho^{-1}f_1(x,D)+Q^4f_2(x,D)\bigg) \ ,
\end{align}
with
\begin{align}
    f_1(x,D)=-\frac{D}{2}(\frac{D}{2}+1)(1-\frac{1}{D})x^{\frac{1}{2}}-\frac{1}{2}(1-\epsilon) \ , \\
    f_2(x,D)=-D(1-\frac{1}{D})x^{\frac{1}{2}}(1+2x^2-2x)-(1-\epsilon)(2x-1)^2+D(1-\frac{1}{D})(1-2x)(1-x^{\frac{1}{2})} \ .
\end{align}
Performing the integrals, one obtains Eq.~(\ref{eq:7nnoncon}).  Similarly, the $T^{ii}_{\rm 9c}$ can be calculated as
\begin{align}\label{eq:5d}
    \langle T^{ii}_{\gamma \perp}\rangle_{\rm 9c} =\frac{ie^2}{m_e}\mu^{2\epsilon}\int \frac{dk^0d^{D}\vec{k}}{(2\pi)^{D+1}}\frac{A(k,Q,D){\rm tr}(P(k-Q)P(k+Q))+2(4-D)Q^TP(k-Q)P(k+Q)Q}{\bigg(k_0^2-(\vec{k}-\vec{Q})^2+i0\bigg)\bigg(k_0^2-(\vec{k}+\vec{Q})^2+i0\bigg)} \ ,
\end{align}
where
\begin{align} \label{eq:AkQD}
    A(k,Q,D)=\frac{D-2}{2}k_0^2+\frac{4-D}{2}( \vec{k}-\vec{Q})\cdot(\vec{k}+\vec{Q}) \ .
\end{align}
After a similar calculation, one has
\begin{align}
    \langle T^{ii}_{\gamma \perp}\rangle_{\rm 9c}=\frac{e^2}{m_e\sqrt{\pi}}(4\pi)^{-\frac{D}{2}}\mu^{2\epsilon}\int_{0}^1 dx\int_{0}^{\infty}d\rho \rho^{-\frac{D}{2}+\frac{1}{2}}e^{-4Q^2\rho x(1-x)}\bigg(\rho^{-1}g_1(x,D)+Q^2g_2(x,D)\bigg) \ ,
\end{align}
with
\begin{align}
    g_1(x,D)=-(1-\epsilon)\bigg[\frac{D}{2}(\frac{1}{2}+\epsilon)-\frac{1}{2}(\frac{1}{2}-\epsilon)\bigg]  \ , \\
    g_2(x,D)=4(1-\epsilon)(\frac{1}{2}+\epsilon)x(1-x)-D(1-2\epsilon)(1-\frac{1}{D})(\sqrt{x}+\sqrt{\bar x}-1) \ .
\end{align}
Performing the integrals, one obtains Eq.~(\ref{eq:7dtrace}).
 \section{Calculation of Fig.~\ref{fig:lambmix}}
 In this appendix we calculate Fig.~\ref{fig:lambmix}. We start with the trace part in Eq.~(\ref{eq:5atrace}). To calculate this one needs (for $E_M>E_N$)
\begin{align}
    \int \frac{dk^0}{2\pi}\frac{2}{k_0^2+\vec{k}^2}\frac{1}{(ik^0-E_M+E_N)(ik^0-0)}=\frac{1}{(|\vec{k}|+E_M-E_N)|\vec{k}|^2} \ ,
\end{align}
therefore by introducing the $\alpha$ and $\lambda$ parameters one has
\begin{align}
\int \frac{dk^0}{2\pi}\frac{2}{k_0^2+\vec{k}^2}\frac{1}{(ik^0-E_M+E_N)(ik^0-0)}
=2\sqrt{\frac{1}{4\pi}}\int_{0}^{\infty}d\alpha\int_{0}^{\infty} \lambda d\lambda\int_{0}^1 dt \sqrt{\alpha}e^{-\alpha\vec{k}^2-\frac{\lambda^2}{4}-\lambda \sqrt{\alpha}t(E_M-E_N)} \ .
\end{align}
Clearly, for $E_N-E_M=0$ it simply reduces to the representation in free NRQED. One then proceeds as usual, which leads to
\begin{align}
    \langle T^{ii}_{\gamma \parallel \perp}\rangle_{\rm 7a}=2(D-2)\frac{e^2}{I_1}Q^2(1-\frac{1}{D})\frac{1}{2}\bigg(\frac{1}{4\pi}\bigg)^{\frac{D}{2}}\mu^{2\epsilon}\sum_{M}i(\vec{v}_{0M}\cdot \vec{x}_{M0}-\vec{x}_{0M}\cdot \vec{v}_{M0})I_{M0} \ ,
\end{align}
where
\begin{align}
    &I_{MN}=\frac{1}{2}\int_{0}^1 dx x^{\frac{1}{2}-\epsilon}\int_{0}^1 dt t^{-2\epsilon} \int_{0}^{\infty} d\lambda \lambda^{1-2\epsilon}e^{-\frac{\lambda^2}{4}}\int_{0}^{\infty} d\rho \rho^{-1+\epsilon} e^{-\sqrt{\rho}}\times (E_M-E_N)^{-2\epsilon} \ ,
\end{align}
can be evaluated easily. Expanding in $\epsilon$,  one obtains the result in Eq.~(\ref{eq:5aresult1}). Similarly, the non-conserved part Eq.~(\ref{eq:5anoncon}), after parameterizing, reads
\begin{align}
    &Q^i \langle T^{ij}_{\gamma \parallel \perp} \rangle_{\rm 7a} Q^j= -Q^4\frac{2e^2}{\sqrt{\pi}}\sum_{M}\frac{2\vec{v}_{M0}\cdot \vec{v}_{0M}}{D(E_M-E_0)}\frac{D}{2}(1-\frac{1}{D})(4\pi)^{-\frac{D}{2}}\nonumber \\
    & \times \int_{0}^1\sqrt{x}(2x-1)dx\int_{0}^1 dt\int_{0}^{\infty}\lambda d\lambda \int_{0}^{\infty}\rho^{-\frac{D}{2}+\frac{1}{2}}d\rho e^{-\frac{\lambda^2}{4}-\lambda\sqrt{\rho x}t(E_M-E_0)} \ .
\end{align}
Evaluating the integrals, one obtains Eq.~(\ref{eq:5anonconresult}).

\section{Calculation of Fig.~\ref{fig:lambradia}}
In this appendix we calculate the pure radiative diagram Fig.~\ref{fig:lambradia}. We first start with the trace part Eq.~(\ref{eq:radiab}). Notice that the pole of the last propagator is located at $k^0=-(E_M-E _0)+i0$, therefore when $E_M>E_0$ one can simply integrate without encountering any poles by $k^0 \rightarrow ik^E$. After this and introducing the $\lambda$ parameter for the eikonal-like propagator, one has
\begin{align}
    \langle T^{ii}_{\gamma \perp}\rangle_{\rm 7c}(2\vec{Q})=\frac{e^2}{\sqrt{\pi}}(4\pi)^{-\frac{D}{2}}\mu^{-2\epsilon}\sum_{M}\frac{\vec{v}_{0M}\cdot\vec{v}_{M0}}{(E_M-E_0)^{1-2\epsilon}}Q^2\bigg[I_{A}\bigg(\frac{Q^2}{(E_M-E_0)^2}\bigg)  +I_B\bigg(\frac{Q^2}{(E_M-E_0)^2}\bigg)+I_C\bigg(\frac{Q^2}{(E_M-E_0)^2}\bigg)\bigg]\ ,
\end{align}
where
\begin{align}
    &I_A(q)=-8(1-\epsilon)(\frac{1}{2}+\epsilon)\int_{0}^{1}dxx\bar x\int_{0}^{\infty}d\lambda\int_{0}^{\infty}\rho^{-\frac{D}{2}+1}d\rho e^{-\frac{\lambda^2}{4}-\lambda\sqrt{\rho}-4\rho x\bar x q} \ , \\
    &I_B(q)=\frac{D}{2}(1-\frac{1}{D})(1-2\epsilon)\int_{0}^1 dx x\bar x\int_{0}^1\frac{dt_1dt_2}{(xt_1+\bar x t_2)}\int_{0}^{\infty} d\lambda\int_{0}^{\infty} \rho^{-\frac{D}{2}+1}d\rho (1-\frac{\lambda^2}{2}) e^{-\frac{\lambda^2}{4}-\lambda \sqrt{\rho}\sqrt{xt_1+\bar x t_2}-4\rho x \bar x q} \ , \\
    &I_C(q)=-\int_{0}^{1} dx x\bar x\int_{0}^1 dt \int_{0}^{\infty} d\lambda \int_{0}^{\infty} \rho^{-\frac{D}{2}+1}d\rho(4+\lambda^2)e^{-\frac{\lambda^2}{4}-\lambda \sqrt{\rho}-4\rho x\bar xt q} \ .
\end{align}
It is easy to check that all the integrals above are absolutely convergent at $D=3$, therefore one can simply set $D=3$ in all the expressions. Furthermore, since we are only interested in the small $q$ asymptotics, we can perform the standard Mellin transform as
\begin{align}
    I_i(s)=\int_{0}^{\infty} dq I_i(q) q^{s-1} \ ,
\end{align}
where $i=A,B,C$. It is easy to show that the Mellin transforms are convergent absolutely for $0<{\rm Re}(s)<\frac{1}{2}$, and $I_i(\sigma+it)$ decay rapidly as $|t| \rightarrow \infty$ for $-1<\sigma<\frac{1}{2}$. Therefore,  according to the well-known relation between asymptotic expansion and Mellin transform,  expansion of $I_i(q)$ at $q=0$ can be easily recovered from the poles of the Mellin transform when ${\rm Re}(s)\le 0$. After simple calculation, one finds
\begin{align}
   &I_A(s)=-\frac{4\Gamma (1-2 s) \Gamma^2(2-s)\Gamma^2 (s)}{\Gamma (4-2 s)} \ , \\
     &I_B(s)=\frac{\sqrt{\pi } \left(4^s (2 s-3)+4\right) \Gamma (1-2 s) \Gamma (1-s) \Gamma^2 (s)}{(2 s+1) \Gamma \left(\frac{5}{2}-s\right)} \ , \\
    &I_C(s)=-\frac{8 \Gamma (s+2)\Gamma^2 (2-s)\Gamma (-2 s) \Gamma(s)}{(s-1) \Gamma (4-2 s)} \ ,
\end{align}
Each of them has double pole at $s=0$, but adding them up, the double pole cancels
\begin{align}
    I_A(s)+I_B(s)+I_C(s)=-\frac{16\ln 4}{3s}+{\cal O}(1) \ ,
\end{align}
and the next pole is at $s=-1$. Therefore, the $I_A+I_B+I_C$ is finite at $Q^2=0$, with value equal to the residue of the Mellin transform at $s=0$. From this one obtains the result Eq.~(\ref{eq:5cresult}).

We then move to the non-conserved part of Fig.~\ref{fig:lambradia}. Using the standard Feynman rule, after certain simplification one has
\begin{align}
    Q^i\langle T^{ij}_{\gamma E}\rangle_{\rm 7c} Q^j= ie^2\sum_{M}\frac{\vec{v}_{0M}\cdot \vec{v}_{M0}}{D}\mu^{2\epsilon}\int \frac{d^{D+1}k}{(2\pi)^{D+1}}\frac{k_0^2\bigg[Q^2{\rm tr}P(k-Q)P(k+Q)-2Q^TP(k-Q)P(k+Q)Q\bigg]}{\bigg(k_0^2-(\vec{k}-\vec{Q})^2+i0\bigg)\bigg(k_0^2-(\vec{k}+\vec{Q})^2+i0\bigg)\bigg(E_0-k^0-E_M+i0\bigg)}\ ,
\end{align}
and
\begin{align}
    Q^i\langle T^{ij}_{\gamma B}\rangle_{\rm 7c} Q^j=ie^2\sum_{M}\frac{\vec{v}_{0M}\cdot \vec{v}_{M0}}{D}\mu^{2\epsilon}\int \frac{d^{D+1}k}{(2\pi)^{D+1}}\frac{{\rm tr}P(k-Q)P(k+Q)A(k,Q)+Q^TP(k-Q)P(k+Q)QB(k,Q)}{\bigg(k_0^2-(\vec{k}-\vec{Q})^2+i0\bigg)\bigg(k_0^2-(\vec{k}+\vec{Q})^2+i0\bigg)\bigg(E_0-k^0-E_M+i0\bigg)} \ ,
\end{align}
where $A(k,Q)$, $B(k,Q)$ are defined before in Eq.~(\ref{eq:ABdef}) and the subscript E, B denotes the electric and magnetic part of the MC density in Eq.~(\ref{eq:TE}) and Eq.~(\ref{eq:TB}), respectively.  By combining them, one has
\begin{align}
    Q^i\langle T^{ij}_{\gamma \perp}\rangle_{\rm 7c} Q^j=e^2\sum_{M}\frac{2\vec{v}_{0M}\cdot \vec{v}_{M0}}{D}F_{M0}
\end{align}
where
\begin{align}
    F_{MN}=&iQ^2\mu^{2\epsilon}\int \frac{d^{D+1}k}{(2\pi)^{D+1}}\frac{1-\epsilon}{\bigg(k_0^2-(\vec{k}+\vec{Q})^2+i0\bigg)\bigg(E_N-E_M-k_0+i0\bigg)}\nonumber \\ &-i\mu^{2\epsilon}\int \frac{d^{D+1}k}{(2\pi)^{D+1}}\frac{\vec{k}^2\vec{Q}^2-(\vec{k}\cdot \vec{Q})^2}{|\vec k|^2|\vec{k}+2\vec{Q}|^2}\frac{\vec{k}^2+2\vec{k}\cdot \vec{Q}+2Q^2}{\bigg(k_0^2-(\vec{k}+\vec{Q})^2+i0\bigg)\bigg(E_N-E_M-k_0+i0\bigg)} \nonumber \\
    &+2i\mu^{2\epsilon}\int \frac{d^{D+1}k}{(2\pi)^{D+1}}\frac{(1-\epsilon)(\vec{k}\cdot \vec{Q})^2}{\bigg(k_0^2-(\vec{k}-\vec{Q})^2+i0\bigg)\bigg(k_0^2-(\vec{k}+\vec{Q})^2+i0\bigg)\bigg(E_N-E_M-k_0+i0\bigg)} \nonumber \\
    &-2i\mu^{2\epsilon}\int \frac{d^{D+1}k}{(2\pi)^{D+1}}\frac{\vec{Q}^2+\vec{k}\cdot \vec{Q}}{\bigg(k_0^2-\vec{k}^2+i0\bigg)\bigg(k_0^2-(\vec{k}+2\vec{Q})^2+i0\bigg)\bigg(E_N-E_M-k_0+i0\bigg)}\frac{\vec{k}^2\vec{Q}^2-(\vec{k}\cdot \vec{Q})^2}{|\vec{k}|^2} \ .
\end{align}
Introducing the parameters, one has
\begin{align}
     F_{MN}=Q^4\mu^{2\epsilon}\frac{(4\pi)^{-\frac{D}{2}}}{\sqrt{\pi}(E_M-E_N)^{1+2\epsilon}}\bigg(F_1+F_2+F_3+F_4+F_5\bigg) \ ,
\end{align}
where $F_i\equiv F_i\bigg(\frac{Q^2}{(E_M-E_N)^2}\bigg)$ are represented as
\begin{align}
    &F_1(q)=-\frac{D}{4}(\frac{D}{2}+1)(1-\frac{1}{D})\int_{0}^1 4x^2\bar x dx\int_{0}^{\infty}d\lambda \int_{0}^1dt_1\int_{0}^1dt_2\int_{0}^{\infty}\rho^{-\frac{D}{2}+1}d\rho e^{-\frac{\lambda^2}{4}-\lambda\sqrt{\rho x t_1}-4\rho x(1-x)t_2 q} \ , \\
    &F_2(q)=\frac{D}{2}(1-\frac{1}{D})\int_{0}^1 x(1+2x^2-2x)dx\int_{0}^{\infty} d\lambda \int_{0}^1 dt_1\int_{0}^{\infty}\rho^{-\frac{D}{2}+1}d\rho e^{-\frac{\lambda^2}{4}-\lambda\sqrt{\rho x t_1}-4\rho x(1-x)q} \ , \\
    &F_3(q)=-\frac{1}{D}\frac{D}{2}(1-\epsilon)\int_{0}^1 4x(1-x)dx\int_{0}^{\infty}d\lambda \int_{0}^1dt_1\int_{0}^{\infty}\rho^{-\frac{D}{2}+1}d\rho e^{-\frac{\lambda^2}{4}-\lambda\sqrt{\rho}-4\rho x(1-x)t_1q} \ , \\ &F_4(q)=(1-\epsilon)\int_{0}^1 (2x-1)^2dx\int_{0}^{\infty}d\lambda\int_{0}^{\infty}\rho^{-\frac{D}{2}+1}d\rho e^{-\frac{\lambda^2}{4}-\lambda\sqrt{\rho}-4\rho x(1-x)q} \ , \\
    &F_5(q)=-\frac{D}{2}(1-\frac{1}{D})\int_{0}^1 x(2x-1)dx\int_{0}^{\infty}d\lambda \int_{0}^1dt_1\int_{0}^{\infty}\rho^{-\frac{D}{2}+1}d\rho e^{-\frac{\lambda^2}{4}-\lambda\sqrt{\rho(xt_1+1-x)}-4\rho x(1-x)q} \ .
\end{align}
Clearly, all the integrals are absolutely convergent for $D=3$ and $q\ne 0$, thus one can set $D=3$ and use the Mellin transform technique as before to obtain the small-$q$ asymptotics. Direct calculation leads to
\begin{align}
    \sum_{i=1}^5 F_i(s)=-\frac{8}{15 s^2}+\frac{8 (15 \ln (4)-46)}{225 s} + {\cal O}(1)\ ,
\end{align}
which implies
\begin{align}
    \sum_{i=1}^5 F_i(q)=\frac{8}{15}\ln q +\frac{8 (15 \ln (4)-46)}{225 }+{\cal O}(q) \ ,
\end{align}
which leads to Eq.~(\ref{eq:5cnonresult}).

\section{Calculation of Eq.~(\ref{eq:finalre})}
In this appendix we estimate the result Eq.~(\ref{eq:finalre}). The intermediate state $M$ must have $l=1$ due to selection rule. For the discrete spectrum, one has the matrix element (in the unit where $m_e=1$ and $\alpha=1$)
\begin{align}
    |\langle n1|x|00\rangle|^2=|\langle n1|y|00\rangle|^2=|\langle n1|z|00\rangle|^2=\bigg(\int\frac{\sqrt{3}}{4\pi}\cos^2 \theta d\cos \theta d\phi\bigg)^2\bigg(\int_{0}^{\infty} dr r^3 R_{00}(r)R_{n1}(r)\bigg)^2 \ .
\end{align}
The radial overlapping turns out to be non-trivial and can be shown to be~\cite{Bethe:1957ncq}
\begin{align}
    \bigg(\int_{0}^{\infty} dr r^3 R_{00}(r)R_{n1}(r)\bigg)^2=\frac{2^8n^7}{(n^2-1)^5}\bigg(1-\frac{2}{n+1}\bigg)^{2n} \ .
\end{align}
Therefore, the matrix element reads
\begin{align}
    \frac{2|\vec{v}_{n1,0}|^2}{3(E_n-E_0)}=\frac{2}{3}|\vec{x}_{n1,0}|^2(E_n-E_0)=\frac{2^8n^5}{3(n^2-1)^4}\bigg(1-\frac{2}{n+1}\bigg)^{2n}  \ .
\end{align}
The discrete spectrum contribution $\tau_{d}$ then reads
\begin{align}
    \tau_{d}=\sum_{n=2}^{\infty}\frac{2^8n^5}{3(n^2-1)^4}\bigg(1-\frac{2}{n+1}\bigg)^{2n}\ln \bigg(1-\frac{1}{n^2}\bigg)^2=-0.264\ .
\end{align}
For the contribution from continuum spectrum, one needs the normalized wave function $R_{El}(r)$ with the normalization condition
\begin{align}
    \int_{0}^{\infty}dr r^2 R_{El}(r)R_{E'l}(r)=\delta(E-E') \ ,
\end{align}
where $E,E'>0$ are the energies of the states. In terms of these, one has
\begin{align}
    \tau_{\rm c}=\frac{1}{3}\int_{0}^{\infty}dE \bigg(\int_{0}^{\infty} dr r^3 R_{00}(r)R_{E1}(r)\bigg)^2 (2E+1) \ln (2E+1)^2\ .
\end{align}
It turns out that the radial overlapping can be worked out explicitly~\cite{Bethe:1957ncq}
\begin{align}
    \bigg(\int_{0}^{\infty} dr r^3 R_{00}(r)R_{E1}(r)\bigg)^2 =\frac{2^8}{(2E+1)^5}\frac{e^{-\frac{4}{\sqrt{2E}}{\rm Arccot}(\frac{1}{\sqrt{2E}})}}{1-e^{-\frac{2\pi}{\sqrt{2E}}}} \ .
\end{align}
Given these, the $\tau_c$ can be evaluated as
\begin{align}
     \tau_{\rm c}=\frac{1}{3}\int_{0}^{\infty}dE \frac{2^8\ln (2E+1)^2}{(2E+1)^4}\frac{e^{-\frac{4}{\sqrt{2E}}{\rm Arccot}(\frac{1}{\sqrt{2E}})}}{1-e^{-\frac{2\pi}{\sqrt{2E}}}}=0.458 \ .
\end{align}
It is different in sign to $\tau_d$, and is about $73$ percent larger. As a consistency check, we have verified that the sum rule Eq.~(\ref{eq:sumrule}) is satisfied numerically with precision of $10^{-10}$.
\section{The QED contribution $\tilde C_{\rm QED}(Q)$}
In this appendix we collect required formulas for QED contribution to the electron's $C$-form factor. According to~\cite{Milton:1977je}, in unit $m_e=1$ the contribution reads
\begin{align}
\tilde C_{\rm QED}(Q^2)= -\frac{e^2 }{4\pi ^2Q^2}\left(\frac{8 \left(1-x^2\right) x^4 F(x)}{\left(x^2+1\right)^5}-\frac{3 x^4+4 x^2+3}{3 \left(x^2+1\right)^2}-\frac{\left(1-x^2\right)^2 \left(x^4+8 x^2+1\right) \ln \frac{1-x^2}{x}}{3 \left(x^2+1\right)^4}-\frac{5 \left(x^2+1\right) \ln x}{6 \left(1-x^2\right)}\right) \ ,
\end{align}
where
\begin{align}
x=\sqrt{\frac{\sqrt{Q^2+1}-Q}{\sqrt{Q^2+1}+Q}} \ , \\
F(x)=-\frac{1}{2}\bigg(\ln^2 x+\frac{\pi^2}{3}-\int_{0}^{x^2} \frac{dt}{t}\ln (1-t)\bigg) \ .
\end{align}
To obtain the small-$Q$ asymptotics, notice that for small $Q$ one has
\begin{align}
    \frac{8 \left(1-x^2\right) x^4 }{\left(x^2+1\right)^5}\frac{1}{Q^2}=\frac{1}{2Q}-\frac{5Q}{4}+{\cal O}(Q^3) \ ,
\end{align}
therefore to obtain the ${\cal O}(Q^0)$ contribution to $\tilde C_{\rm QED}(Q^2)$ one only needs to expand $F(x)$ to linear order in $Q$, which can be done by
\begin{align}
    -\int_{0}^{x^2} \frac{dt}{t}\ln (1-t)=-\int_{0}^{1-2Q+{\cal O}(Q^2)} \frac{dt}{t}\ln (1-t)=\frac{\pi^2}{6}+2Q(\ln 2Q-1)+{\cal O}(Q^2) \ .
\end{align}
Using these relations, one obtains the small-$Q$ expansion
\begin{align}
    \tilde C_{\rm QED}(Q^2)=\frac{e^2}{32 Q}+\frac{e^2}{6 \pi ^2} \ln 4Q^2-\frac{11e^2}{72 \pi ^2}+{\cal O}(Q) \ ,
\end{align}
which is nothing but Eq.~(\ref{eq:CQED}), after restoring the $m_e$ dependency.
\end{widetext}

\bibliography{bibliography}

\end{document}